\documentclass[twocolumn,tighten,trackchanges,twocolappendix]{aastex701}
\usepackage{CJK}
\usepackage{amsmath}
\usepackage{multirow}

\usepackage{xspace}

\newcommand{\Mch}{M_{\rm Ch}}
\newcommand{\Msun}{M_{\rm \odot}}
\newcommand{\Ni}{$^{56}$Ni\xspace}

\defcitealias{millerZTFEarlyObservations2020}{M20}
\defcitealias{PaperI}{Paper I}
\defcitealias{PaperII}{Paper II}

\begin{document}
% share link: https://www.overleaf.com/2646314997hwxbqrcvwqzx#b22650
\title{Decoding the Early-Time Light Curves of Type Ia Supernovae. II. Population Parameters of One Thousand ZTF Supernovae}

\author[0000-0002-7866-4531,gname=Chang,sname=Liu]{Chang~Liu \begin{CJK*}{UTF8}{gbsn}(刘畅)\end{CJK*}}
\affil{Department of Physics and Astronomy, Northwestern University, 2145 Sheridan Rd, Evanston, IL 60208, USA}
\affil{Center for Interdisciplinary Exploration and Research in Astrophysics (CIERA), Northwestern University, 1800 Sherman Ave, Evanston, IL 60201, USA}
\affil{NSF-Simons AI Institute for the Sky (SkAI), 172 E. Chestnut St., Chicago, IL 60611, USA}
\email{ptg.cliu@u.northwestern.edu}

\author[0000-0001-9515-478X]{Adam~A.~Miller}
\affil{Department of Physics and Astronomy, Northwestern University, 2145 Sheridan Rd, Evanston, IL 60208, USA}
\affil{Center for Interdisciplinary Exploration and Research in Astrophysics (CIERA), Northwestern University, 1800 Sherman Ave, Evanston, IL 60201, USA}
\affil{NSF-Simons AI Institute for the Sky (SkAI), 172 E. Chestnut St., Chicago, IL 60611, USA}
\email{amiller@northwestern.edu}

% Main collaborators
\author[0000-0003-2700-1030]{Nikhil~Sarin}
\affil{Kavli Institute for Cosmology, University of Cambridge, Madingley Road, CB3 0HA, UK}
\affil{Institute of Astronomy, University of Cambridge, Madingley Road, CB3 0HA, UK}
\email{nsarin.astro@gmail.com}

\author[0000-0002-9770-3508]{Kate~Maguire}
\affil{School of Physics, Trinity College Dublin, The University of Dublin, Dublin 2, Ireland}
\email{kate.maguire@tcd.ie}

\author[0000-0003-3939-7167]{Tom\'{a}s~E.~M\"{u}ller-Bravo}
\affil{School of Physics, Trinity College Dublin, The University of Dublin, Dublin 2, Ireland}
\affil{Instituto de Ciencias Exactas y Naturales (ICEN), Universidad Arturo Prat, Chile}
\email{t.e.muller-bravo@tcd.ie}

% Other contributors
\author[0000-0003-0853-6427]{Ping~Chen}
\affil{Institute for Advanced Study in Physics, Zhejiang University, Hangzhou 310058, China}
\affil{Institute for Astronomy, School of Physics, Zhejiang University, Hangzhou 310058, China}
\email{ping.chen@zju.edu.cn}

\author[0000-0002-1296-6887]{L.~Galbany}
\affil{Institute of Space Sciences (ICE-CSIC), Campus UAB, Carrer de Can Magrans, s/n, E-08193 Barcelona, Spain.}
\affil{Institut d'Estudis Espacials de Catalunya (IEEC), 08860 Castelldefels (Barcelona), Spain}
\email{lgalbany@ice.csic.es}

\author[0000-0002-1031-0796]{Young-Lo~Kim}
\affil{Department of Astronomy \& Center for Galaxy Evolution Research, Yonsei University, Seoul 03722, Republic of Korea}
\email{olo8308@gmail.com}

\author[0000-0003-1546-6615]{Jesper~Sollerman}
\affil{The Oskar Klein Centre, Department of Astronomy, Stockholm University, Albanova University Center, SE-106 91 Stockholm, Sweden}
\email{jesper@astro.su.se}

% Builders
\author[0000-0001-8018-5348]{Eric~C.~Bellm}
\affiliation{DIRAC Institute, Department of Astronomy, University of Washington, 3910 15th Avenue NE, Seattle, WA 98195, USA}
\email{ecbellm@uw.edu}

\author[0000-0002-0987-3372]{Joahan~Castaneda~Jaimes}
\affil{Division of Physics, Mathematics and Astronomy, California Institute of Technology, 1200 E. California Blvd, Pasadena, CA 91125, USA}
\email{jocastan@ipac.caltech.edu}

\author[0000-0001-9152-6224]{Tracy~X.~Chen}
\affil{IPAC, California Institute of Technology, 1200 E. California Blvd, Pasadena, CA 91125, USA}
\email{xchen@ipac.caltech.edu}

\author[0000-0002-3168-0139]{Matthew~J.~Graham}
\affil{Division of Physics, Mathematics and Astronomy, California Institute of Technology, Pasadena, CA 91125, USA}
\email{mjg@caltech.edu}

\author{David~Hale}
\affil{Caltech Optical Observatories, California Institute of Technology, Pasadena, CA 91125, USA}
\email{dhale@caltech.edu}

\author[0000-0002-5619-4938]{Mansi~M.~Kasliwal}
\affil{Division of Physics, Mathematics and Astronomy, California Institute of Technology, Pasadena, CA 91125, USA}
\email{mansi@astro.caltech.edu}

\author[0000-0003-1227-3738]{Josiah~Purdum}
\affil{Caltech Optical Observatories, California Institute of Technology, Pasadena, CA 91125, USA}
\email{jpurdum@caltech.edu}

\correspondingauthor{Chang~Liu}
\email{ptg.cliu@u.northwestern.edu}


\begin{abstract}
Early-time light curves of Type Ia Supernovae (SNe\,Ia) encode critical information about their progenitor systems. We characterize the rise of normal SNe\,Ia using a volume-complete sample of 972 events from the Zwicky Transient Facility Data Release 2, an order of magnitude larger than any previous dataset for similar analyses. Fitting light curves up to 30\% of peak flux with a power-law model under a hierarchical Bayesian framework, we provide robust population-level constraints on the rise time ($t_\mathrm{rise}$; $\mu=18.55\pm0.08$\,days, $\sigma=1.42\pm0.07$\,days), rise index ($\alpha$; $\mu=2.10\pm0.04$, $\sigma=0.48\pm0.03$ in ZTF $r$), and $g-r$ color evolution ($\alpha_g - \alpha_r$; $\mu=0.20\pm0.02$, $\sigma=0.17\pm0.02$).
These power-law fits are sensitive to the chosen truncation epoch if data beyond $\sim$40\% of peak flux are included, but generally converge when restricted to earlier epochs. 
The relation between rise morphology and light-curve width (\texttt{SALT2} $x_1$ stretch) bifurcates into two distinct regimes: high-stretch SNe\,Ia show clear trends where a higher $x_1$ correlates with shallower rises and more persistent blue colors, whereas low-stretch SNe\,Ia lack such trends. While rise times correlate positively with $x_1$ overall, this relation flattens significantly within the high-stretch population.
Searching for anomalies, we identify several normal SNe\,Ia with unusually long rise times, which potentially exhibit short-duration ($\lesssim$2\,days) flux excesses over a smooth rise. Long-duration ($\sim$5\,days) flux excesses appear common within the high-stretch population and are tied to the shallow rises and early blue colors, pointing to widespread outward $^{56}$Ni mixing.
Multi-dimensional explosion models with more realistic progenitor setups are needed to fully reproduce the observed dichotomy in rise morphology and stretch.
\end{abstract}



\section{Introduction} \label{sec:intro}
Modern time-domain surveys have transformed Type Ia supernovae (SNe\,Ia) from individual curiosities into rich statistical samples, revealing a demographic complexity that defies a single progenitor channel for the thermonuclear explosion of a carbon-oxygen white dwarf (WD) in a binary system \citep[see][for extensive reviews]{liuTypeIaSupernova2023,ruiterTypeIaSupernova2025}.
Although the remarkably standardizable light curves of SNe\,Ia near maximum luminosity have cemented their role as precise cosmological distance indicators \citep{riessObservationalEvidenceSupernovae1998,perlmutterMeasurements42HighRedshift1999,abbottDarkEnergySurvey2024}, this standardization obscures the diverse physical mechanisms driving their explosions.

The early-time light curves ($\lesssim$1 week after the explosion) of SNe\,Ia provide a unique opportunity to break the degeneracy between different explosion channels and progenitor properties. Observationally, early emission is far more diverse in both morphology \citep{firthRisingLightCurves2015,zhengEmpiricalFittingMethod2017a,millerZTFEarlyObservations2020} and color evolution \citep{stritzingerRedBlueEarly2018,bullaZTFEarlyObservations2020,niInfantTypeIa2025} than the light curves near peak brightness. Physically, early-time emission is highly sensitive to the distribution of \Ni in the outer ejecta \citep{piroWHATCANWE2013,piroCONSTRAINTSSHALLOW562014,mageeModellingEarlyTime2018,mageeDetermining56Ni2020}, offering signatures of the underlying explosion mechanism. 

Beyond uncovering the structure in the outermost ejecta, early-time observations are essential for anchoring the explosion epoch and measuring the total rise time to peak luminosity: governed by the diffusion timescale of photons from radioactive \Ni decay, the rise time is essential for estimating the total ejecta mass and \Ni yield \citep{arnettTypeSupernovaeAnalytic1982}. Previous studies have inferred a broad progenitor mass distribution characterized by a substantial fraction of sub-$\Mch$ explosions \citep{scalzoTypeIaSupernova2014,scalzoProbingTypeIa2019,sarinLightcurveModelling22052026}. However, due to the historical scarcity of prompt photometric observations, many of these studies relied on empirical rise-decline correlations derived from small samples \citep[e.g.,][]{ganeshalingamRisetimeDistributionNearby2011}, potentially biasing mass calculations. 

Early-time light curves also uniquely encode signatures of external interactions, where the expanding SN ejecta collides with either a non-degenerate companion \citep{kasenSEEINGCOLLISIONSUPERNOVA2009} or circumstellar material \citep[CSM;][]{piroEXPLORINGPOTENTIALDIVERSITY2016} to produce short-lived flux excesses. Early excesses have been extensively searched for in SNe\,Ia light curves \citep[e.g.,][]{haydenSINGLEDOUBLEDEGENERATE2010,ollingNoSignatureEjecta2015,jiangSurfaceRadioactivityInteractions2018,jiangHSCSSPTransientSurvey2020,deckersConstrainingTypeIa2022,mageeDetectionEfficiencyType2022,fausnaughFourYearsType2023,yeSearchingBumpsCosmological2024,wuWFSTSupernovaeFirst2026,rojas-bravoNotAllWho2026} and confidently found in a handful of normal events \citep[e.g.,][]{zhengVERYYOUNGTYPE2013, marionSN2012cgEVIDENCE2016,hosseinzadehEarlyBlueExcess2017, millerEarlyObservationsType2018,dimitriadisK2ObservationsSN2018,liPhotometricSpectroscopicProperties2018,ashallSpeedBumpSN2022,saiObservationsVeryYoung2022,wangFlightBumblebeeEarly2024,iskandarSN2021hprNormal2025}, providing rare evidence for the binary companion and the mass-transfer history of the progenitor system. Nevertheless, most of these studies remain inconclusive regarding the physical origin of the excesses.

Historically, early-time SN\,Ia studies focused on inferring a global baseline model from aggregate data, defining a ``standard'' event through its mean rise time and strict rise-fall correlations \citep[e.g.,][see \citealp{PaperI}, hereafter \citetalias{PaperI}, for a comprehensive review]{riessRiseTimeNearby1999,goldhaberTimescaleStretchParameterization2001,conleyRiseTimeType2006,ganeshalingamRisetimeDistributionNearby2011,gonzalez-gaitanRISETIMENORMAL2011}. Modern analyses of large, uniform datasets have shifted this paradigm, pulling the focus towards rigorously quantifying the intrinsic scatter around these universal templates.
By modeling the rises of individual SNe from large samples collected from untargeted, all-sky surveys, a significant intrinsic scatter in the rise parameters has been quantified \citep[e.g.,][]{firthRisingLightCurves2015,zhengEmpiricalFittingMethod2017a,millerZTFEarlyObservations2020}. However, aggregating fitting results directly from individual SNe, if not conducted carefully, can introduce dramatic systematics. 
For instance, \citeauthor{millerZTFEarlyObservations2020} (\citeyear{millerZTFEarlyObservations2020}; hereafter \citetalias{millerZTFEarlyObservations2020}) identified a spurious correlation between the inferred rise time and the redshift of the SN, which they attributed to Malmquist bias, noting that higher-redshift SNe are more likely to be discovered at later phases. 
While limiting the analysis to a small subset of well-sampled SNe can mitigate this bias, it also significantly reduces the sample size and may introduce selection biases (e.g., preferentially selecting SNe with longer rise times).

To overcome these challenges, \citetalias{PaperI} developed a hierarchical Bayesian framework that models a bulk SN sample without dedicated early-time data quality cuts, allowing for the robust inference of population-level properties while properly accounting for measurement uncertainty and sampling cadence. In this second paper of the series, we apply this framework to a volume-complete SN\,Ia sample from the Zwicky Transient Facility \citep{bellmZwickyTransientFacility2018,grahamZwickyTransientFacility2019,dekanyZwickyTransientFacility2020} Data Release 2\footnote{Here and hereafter, ``ZTF DR2'' denotes the dedicated ZTF SN\,Ia Data Release 2, distinct from the broader ZTF Public Data Release 2.} \citep[ZTF DR2;][]{rigaultZTFSNIa2025,amenoucheZTFSNIa2025}. After outlining our model implementation (Section~\ref{sec:setup}), we derive early-time SN\,Ia population demographics from the ZTF DR2 and evaluate the impact of data quality cuts (Section~\ref{sec:ztf}). In Section~\ref{sec:corr_x1}, we link these early-time observables to the full light-curve stretch parameter ($x_1$) from \texttt{SALT2} \citep{guySALT2UsingDistant2007}. Finally, we investigate SNe with anomalous early-time behavior, compare our results to theoretical near- and sub-$\Mch$ models (Section~\ref{sec:discussion}), and conclude in Section~\ref{sec:conclusions}.

\section{Model Setup} \label{sec:setup}

\begin{deluxetable}{lcll}
\tablenum{1}
\tablecaption{Prior Distributions of Model Parameters \label{tab:priors}}
\tablehead{
\colhead{Parameter} & \colhead{Unit} & \colhead{Description} & \colhead{Prior}
}
\startdata
$C_d$ & flux unit\tablenotemark{a} & Baseline flux in fcqfID $d$ & Uniform($-50, 50$) \\
$\mu_{t_\mathrm{rise}}$ & day & Mean rise time of the population & Uniform($5, 35$) \\
$\mu_{\alpha_b}$ & \nodata & Mean power-law index of the population in filter $b$ & Uniform($1, 4$) \\
$\mu_{\ln A_b}$ & \nodata & Mean log-amplitude of the population in filter $b$ & Uniform($0, \ln(10^{3})$) \\
$\sigma_{t_\mathrm{rise}}$ & day & Scatter in rise time of the population & HalfCauchy($1.5$) \\
$\sigma_{\alpha_b}$ & \nodata & Scatter in power-law index of the population in filter $b$ & HalfCauchy($0.3$) \\
$\sigma_{\ln A_b}$ & \nodata & Scatter in log-amplitude  of the population in filter $b$ & HalfCauchy($0.5$) \\
$\mathbf{L}$ & \nodata & Cholesky decomposition of correlation matrix & LKJ(1) \\
\enddata
\tablenotetext{a}{The flux has been normalized such that 100 flux units correspond to the peak brightness of the SN in the respective filter.}
\end{deluxetable}

We model the early-time SN\,Ia light curves using an empirical power-law rise \citepalias{millerZTFEarlyObservations2020}. For each SN $k$, the flux $f_{b,k}(t)$ in filter $b$ is parameterized by the time of first light ($t_{\mathrm{fl}, k}$, assumed to be identical across all filters), a filter-specific amplitude ($A_{b,k}$), and a power-law rise index ($\alpha_{b,k}$). We also incorporate a constant baseline flux ($C_d$) to account for residual flux in the reference image prior to SN discovery for each specific field, CCD quadrant, and filter\footnote{In \citetalias{PaperI}, $C$ is assumed to be only SN- and filter-specific. This represents a special case where all observations of an SN in a given filter are obtained from the same field and CCD quadrant.} \citep[fcqfID $d$;][]{yaoZTFEarlyObservations2019}:
\begin{equation}
    f_{b,k}(t) = C_d + H(t-t_{\mathrm{fl}, k}) A_{b,k} \left(\frac{t-t_{\mathrm{fl}, k}}{t_\mathrm{pivot}}\right)^{\alpha_{b,k}}.
\end{equation}
Here, $H$ is the Heaviside step function, ensuring that the supernova flux is zero before $t_\mathrm{fl}$, and $t_\mathrm{pivot}=8$\,days is a fixed epoch chosen to minimize the covariance between $A$ and $\alpha$, thereby boosting sampling efficiency \citepalias{PaperI}.
The observed flux is normalized by the peak flux in each respective filter. To ensure that the power-law approximation remains valid, we restrict our fits to an early-time truncation threshold, when the SN flux reaches a certain fraction of the peak flux (e.g., 30\%).

We adopt the hierarchical Bayesian framework established in \citetalias{PaperI}, which simultaneously constrains individual and population-level parameters. At the individual level, each SN has its own rise time $t_\mathrm{rise}$, filter-specific power-law rise index $\alpha$, and amplitude $\ln A$. At the population level, these parameters are drawn from a multivariate Gaussian distribution (the population prior), whose mean, scatter, and covariance structure are themselves constrained simultaneously with the individual fits. This joint inference naturally re-weights each SN's contribution to the population-level result according to its data quality, allowing sparsely sampled events and noisy detections to be included without bias. As demonstrated in \citetalias{PaperI}, the hierarchical approach dramatically outperforms the traditional two-step method of fitting individual SNe and then aggregating the results, robustly recovering population-level parameters free of spurious scatter and correlations. The priors are listed in Table~\ref{tab:priors}.

For real observations, the peak time $t_0$ is inferred from the full light curve (here we adopt the \texttt{SALT2} fits from \citet{rigaultZTFSNIa2025a}), which introduces additional uncertainty into the rise time $t_\mathrm{rise} = t_0 - t_\mathrm{fl}$. Additionally, the ZTF image-differencing pipeline may underestimate photometric uncertainties \citep{yaoZTFEarlyObservations2019}, potentially biasing the inference of early-time parameters. To account for these effects, in the likelihood function:
\begin{equation}
    \ln \mathcal{L} = -\frac{1}{2}\sum_{i}\left[
    \frac{(f_{k, d}(t_i) - f_{\mathrm{obs}, i})^2}{\sigma_{i}^2} + \ln(2\pi\sigma_{i}^2)
    \right],
\end{equation}
where the summation is over all observations $i$ for SN $k$ in fcqfID $d$ (we omit the SN and fcqfID indices hereafter for clarity); $f_{k, d}(t_i)$ is the model flux at time $t_i$, and $f_{\mathrm{obs},i}$ is the corresponding observed flux. We define a modified uncertainty term $\sigma_{i}$. 
\begin{equation}
    \sigma_{i}^2 = (\beta_d\sigma_{\mathrm{obs}, i})^2 + \sigma_{f(t_\mathrm{fl}),i}^2,
\end{equation}
where $\sigma_{\mathrm{obs}, i}$ is the reported flux uncertainty. The scaling factor $\beta_d \ge 1$ corrects for underestimated photometric uncertainties in each fcqfID $d$, calibrated using pre- and post-SN baseline observations in ZTF DR2.
\footnote{We do not apply the error floor proposed by \citet{amenoucheZTFSNIa2025}, which calibrates light-curve residuals against SALT2 fits. Our fitting process does not rely on SALT2 models, and $\beta_d$ already provides an empirical correction for underestimated uncertainties.} 
Furthermore, the additional term $\sigma_{f(t_\mathrm{fl}), i}$ translates the uncertainty in $t_{0, k}$ to a flux uncertainty at time $t_i$:
\begin{equation}
    \sigma_{f(t_\mathrm{fl}), i} = \sigma_{t_{0, k}}\left|\frac{\mathrm d}{\mathrm dt_{\mathrm{fl}, k}} f_{k,d}(t;t_\mathrm{fl})\right|_{t_i},
\end{equation}
which is evaluated analytically for our power-law model.

We implement the model in \texttt{NumPyro} \citep{binghamPyro2019, phanComposableEffectsFlexible2019} and use the No-U-Turn Sampler \citep[NUTS;][]{hoffmanNoUTurnSamplerAdaptively2014} to draw samples from the posterior distribution. 
To ensure robust convergence of the NUTS sampler, we employ a multi-stage initialization strategy. First, we construct a simulated annealing phase with gradual likelihood tempering, slowly transitioning from the prior to the full likelihood.
Then we initialize the subsequent NUTS sampler using the median parameter values from the final 10--20\% of the annealing iterations.
The additional conditional dependence introduced by $\sigma_{t_0}$ complicates the posterior geometry, hindering exploration during the early sampling phase. To mitigate this, we fix $\sigma_{t_0}=0$ during the initial 25\% of warm-up steps. We then re-initialize the sampler using the median posterior values from these $\sigma_{t_0}$-free samples, and proceed with the remaining 75\% of warm-up steps and the main sampling phase with the true $\sigma_{t_0}$ value. We use 3,000 total warm-up steps as default, and increase the steps if the sampled chains have not fully converged.

We run 12 vectorized chains on NVIDIA (A30/L40S) GPUs.
We check for convergence using the Gelman-Rubin statistic $\hat R$ and effective sample size $N_\mathrm{eff}$: we require $\hat R \le 1.01$ and $N_\mathrm{eff} \ge 10^3$ for all random variables to ensure good convergence.

\section{Application on ZTF DR2} \label{sec:ztf}

\begin{figure*}
    \includegraphics[width=\linewidth]{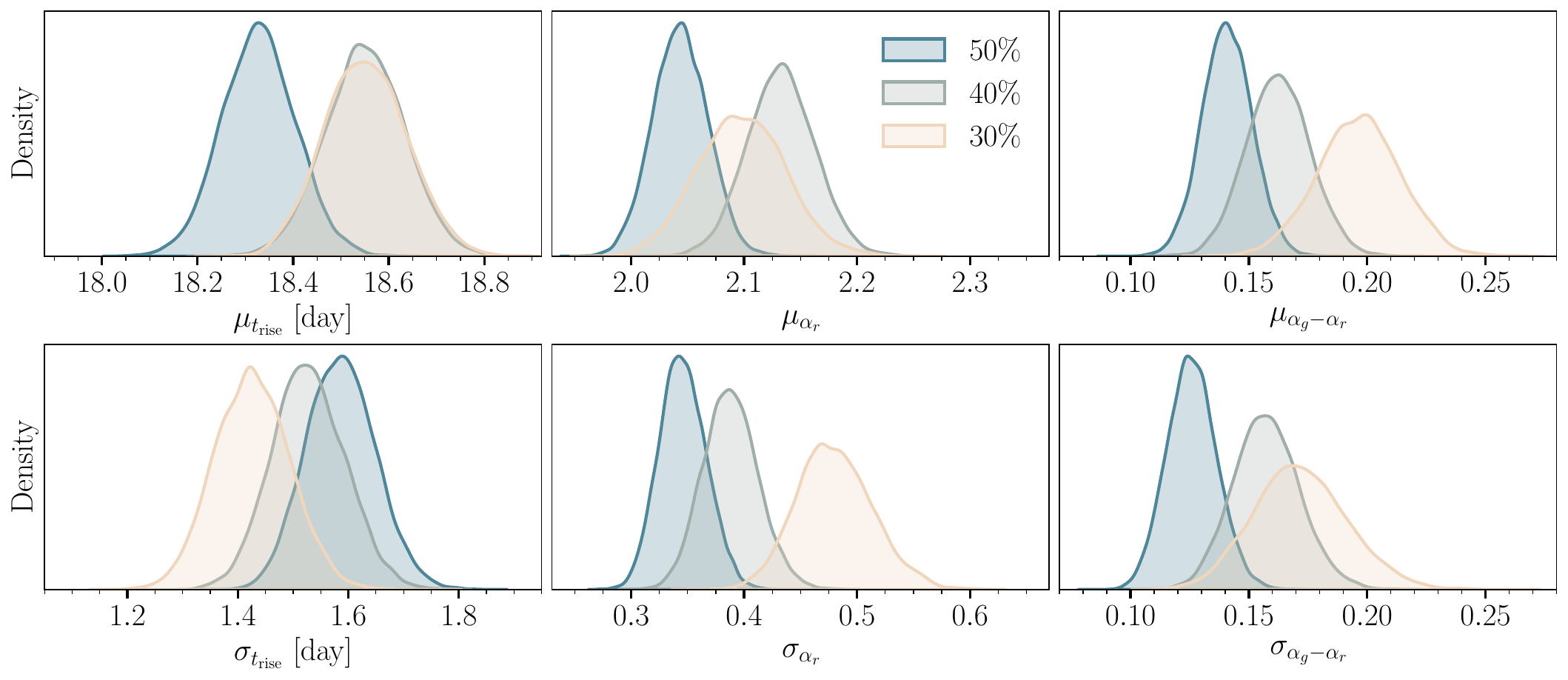}
    \caption{Population-level mean and scatter of early-time parameters in the ZTF DR2 volume-complete sample inferred with the hierarchical Bayesian framework. The posterior distributions of the mean and scatter for the rise time ($t_\mathrm{rise}$), the $r$-band rise index ($\alpha_r$), and the early-time color evolution index ($\alpha_g - \alpha_r$) are shown for three different truncation thresholds: 50\%, 40\%, and 30\%.}
    \label{fig:ztf_mean_scatter}
\end{figure*}

\begin{figure*}
    \centering
    \includegraphics[width=\linewidth]{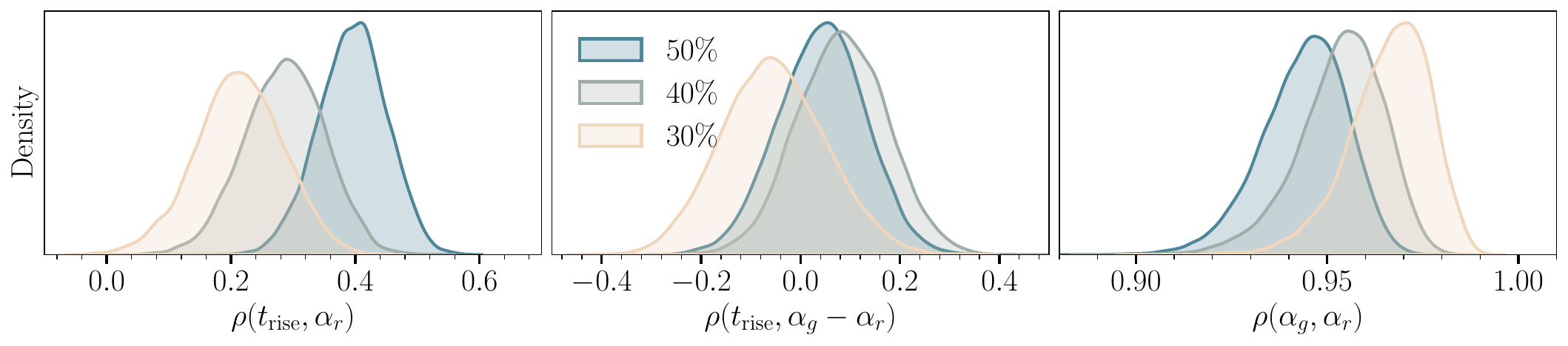}
    \caption{Population-level parameter correlations in the ZTF DR2 volume-complete sample. The posterior distributions of the correlation coefficients between early-time parameters are shown across different flux truncation thresholds. We recover a near-perfect correlation between the $g$- and $r$-band rise indices and a moderate positive correlation between $t_\mathrm{rise}$ and $\alpha_r$. No significant correlation is found between $t_\mathrm{rise}$ and the color evolution index ($\alpha_g - \alpha_r$).}
    \label{fig:ztf_corr}
\end{figure*}

\begin{deluxetable*}{lcccc}
\tablecaption{Inferred Population Parameters for Early-Time ZTF SN\,Ia Light Curves \label{tab:hierarchical_params}}
\tablewidth{0pt}
\tablehead{
\colhead{Parameter} & \colhead{Unit} & \colhead{50\% Threshold} & \colhead{40\% Threshold} & \colhead{30\% Threshold}
}
\startdata
\multicolumn{5}{c}{\textbf{Population Means}} \\
$\mu_{t_{\rm rise}}$ & day & $18.33^{+0.08}_{-0.08}$ & $18.55^{+0.08}_{-0.08}$ & $18.55^{+0.09}_{-0.08}$ \\
$\mu_{\alpha_g}$ & \nodata & $2.184^{+0.026}_{-0.025}$ & $2.295^{+0.033}_{-0.034}$ & $2.294^{+0.043}_{-0.042}$ \\
$\mu_{\alpha_r}$ & \nodata & $2.043^{+0.024}_{-0.024}$ & $2.133^{+0.030}_{-0.030}$ & $2.097^{+0.038}_{-0.039}$ \\
$\mu_{\alpha_g-\alpha_r}$ & \nodata & $0.141^{+0.010}_{-0.010}$ & $0.162^{+0.013}_{-0.013}$ & $0.197^{+0.017}_{-0.017}$ \\
[1ex]
\multicolumn{5}{c}{\textbf{Population Scatters}} \\
$\sigma_{t_{\rm rise}}$ & day & $1.59^{+0.06}_{-0.06}$ & $1.53^{+0.07}_{-0.06}$ & $1.42^{+0.07}_{-0.07}$ \\
$\sigma_{\alpha_g}$ & \nodata & $0.380^{+0.022}_{-0.021}$ & $0.475^{+0.027}_{-0.026}$ & $0.583^{+0.036}_{-0.035}$ \\
$\sigma_{\alpha_r}$ & \nodata & $0.345^{+0.021}_{-0.020}$ & $0.388^{+0.025}_{-0.024}$ & $0.479^{+0.035}_{-0.031}$ \\
$\sigma_{\alpha_g-\alpha_r}$ & \nodata & $0.126^{+0.011}_{-0.010}$ & $0.157^{+0.014}_{-0.014}$ & $0.171^{+0.020}_{-0.019}$ \\
[1ex]
\multicolumn{5}{c}{\textbf{Correlation Coefficients}} \\
$\rho(t_{\rm rise}, \alpha_g)$ & \nodata & $0.37^{+0.05}_{-0.06}$ & $0.26^{+0.06}_{-0.06}$ & $0.16^{+0.07}_{-0.07}$ \\
$\rho(t_{\rm rise}, \alpha_r)$ & \nodata & $0.40^{+0.05}_{-0.06}$ & $0.29^{+0.06}_{-0.07}$ & $0.21^{+0.07}_{-0.07}$ \\
$\rho(t_{\rm rise}, \alpha_g-\alpha_r)$ & \nodata & $0.04^{+0.08}_{-0.09}$ & $0.08^{+0.09}_{-0.09}$ & $-0.06^{+0.11}_{-0.10}$ \\
$\rho(\alpha_g, \alpha_r)$ & \nodata & $0.94^{+0.01}_{-0.01}$ & $0.95^{+0.01}_{-0.01}$ & $0.97^{+0.01}_{-0.01}$ \\
\enddata
\tablecomments{Posterior medians and 16th--84th percentile intervals for the rise times, rising slopes, and the color evolution index. Results are shown as a function of the truncation threshold used to define the early-time phase ($50\%$, $40\%$, and $30\%$ of peak flux).}
\end{deluxetable*}

ZTF DR2 provides forced-photometry light curves derived from difference images output by the IPAC processing pipeline \citep{masciZwickyTransientFacility2018}. The detailed data reduction is described in \citet{dr2_phot}.
The full ZTF DR2 contains 3,826 spectroscopically confirmed SNe\,Ia. Of these, 2,667 pass the cosmology-ready quality cuts defined by \citet{rigaultZTFSNIa2025}, with 2,629 classified as spectroscopically normal SNe\,Ia. Here the overluminous 91T-like SNe\,Ia \citep{filippenkoPeculiarTypeIA1992} are also considered normal. To ensure statistical completeness, we restrict our sample to $z < 0.06$ \citep{amenoucheZTFSNIa2025}, yielding a final volume-complete sample of 972 normal objects. The key assumption we make is that this volume-complete sample accurately represents the underlying normal SN\,Ia population, whose early-time parameters can be described by a unimodal, multivariate Gaussian distribution. 

In our primary analysis, we refrain from applying further quality cuts based on early-time data coverage, retaining even those SNe with sparse sampling. Crucially, the hierarchical Bayesian framework naturally down-weights events with a limited number of detections, while still leveraging their constraining power to the population-level parameters. 
Finally, we neglect $K$-corrections in this analysis despite the sample extending to $z=0.06$. As demonstrated in Appendix~\ref{sec:mock_kcorr}, ignoring $K$-corrections introduces only minor systematics to the inferred population-level parameters, which remain well within typical measurement uncertainties.

We apply the standard quality cuts described by \citet{rigaultZTFSNIa2025} to individual photometric measurements, removing any observations with bits 1, 2, 4, 8, or 16 set in their \texttt{flag} attribute. Additionally, we exclude data points in the baseline (taken $>$$25$\,days prior to the \texttt{SALT2} peak time $t_0$) with an apparent signal-to-noise ratio (S/N) above 10, as these typically indicate significant, improperly subtracted baseline offsets. Furthermore, we identify three SNe where $>$$50$\% of the baseline measurements within a single fcqfID are flagged as outliers; for these events, we exclude all observations from the affected fcqfID.

We simultaneously fit all the remaining measurements up to three different truncation thresholds (50\%, 40\%, and 30\% of the peak flux) to evaluate the impact of potential model misspecification for the simple power-law rise.
Ultimately there are $\sim$$10^5$ measurements from $-100$\,days to the epoch of the truncation threshold that pass these quality cuts. Roughly 20 SNe have no measurements before the truncation threshold, and are effectively excluded from the analysis.

Table~\ref{tab:hierarchical_params} outlines the inferred population-level parameters for the early-time ZTF SN\,Ia light curves. Figure~\ref{fig:ztf_mean_scatter} displays the corresponding population mean and scatter for $t_\mathrm{rise}$, $\alpha_r$, and $\alpha_g-\alpha_r$ across various peak flux truncation thresholds (50\%, 40\%, and 30\%). Reducing this threshold from 50\% to 40\% results in systematically higher values for both $\mu_{t_\mathrm{rise}}$ and $\mu_{\alpha_r}$. Below the 40\% threshold, these parameters stabilize: $\mu_{t_\mathrm{rise}}$ converges robustly to 18.55\,days, while the mean rise indices $\mu_{\alpha_g}$ and $\mu_{\alpha_r}$ settle near 2.3 and 2.1, respectively, suggesting systematically steeper rises compared to the fireball model ($\alpha=2$). The mean rise time is broadly consistent with prior studies that modeled individual SN light curves. For instance, both \citet{firthRisingLightCurves2015} and \citetalias{millerZTFEarlyObservations2020} reported a mean rise time of $\sim$19\,days for normal SNe\,Ia, while \citet{haydenRISEFALLTYPE2010} and \citet{zhengEmpiricalFittingMethod2017a} found slightly shorter mean rise times of $\sim$17\,days. \citet{fausnaughFourYearsType2023} reported an even shorter mean of 15.7\,days from a sample collected with the Transiting Exoplanet Survey Satellite (TESS). Upon visual inspection of the intersection between the TESS and ZTF DR2 samples,\footnote{We exclude events for which TESS did not observe the peak flux or the inferred first-light time falls in data gaps, leaving six SNe shared by both samples.} the ZTF flux is always non-zero at the TESS-estimated first-light epoch,\footnote{The only exception is SN2019wvr. For this event, the epoch of maximum light differs by 30\,days in TESS and ZTF, and the TESS light curve is highly noisy and unreliable.} suggesting that the TESS fits are limited by the shallow depth. However, direct comparisons remain challenging, as none of these previous studies utilized a volume-complete sample with a statistically significant size.

Concurrently, the scatter in all light-curve parameters is statistically significant. The rise-time scatter ($\sigma_{t_\mathrm{rise}}$) mildly decreases when limiting the fit to earlier epochs, reaching $\sim$1.4\,days at a truncation threshold of 30\%. The scatter in the $r$-band rise index ($\sigma_{\alpha_r}$) broadens substantially when lowering the threshold from 40\% to 30\%, despite $\mu_{\alpha_r}$ remaining largely stable. This behavior implies that the intrinsic diversity in rise indices narrows as the SNe evolve toward later phases.

The mean color evolution index ($\mu_{\alpha_g - \alpha_r}$) has a positive value of $\sim$0.14--0.20, indicating that the early-time $g$-band flux evolves more steeply than the $r$-band flux. This suggests that, on average, SNe\,Ia become bluer at early times, consistent with previous findings \citep{stritzingerRedBlueEarly2018,bullaZTFEarlyObservations2020,burkeEarlyLightcurvesType2025}, although there is significant scatter within the population ($\sigma_{\alpha_g - \alpha_r}\simeq0.13$--0.17).

In Figure~\ref{fig:ztf_corr}, we present the inferred correlations between the power-law parameters. The posterior distributions of the correlation coefficients across different flux thresholds agree within the $1\sigma$ level. We observe a moderate positive correlation between $t_\mathrm{rise}$ and $\alpha_r$, which weakens ($\rho$ decreases from 0.4 to 0.2) as the truncation threshold is lowered from $50\%$ to $30\%$. While a positive correlation was previously reported in \citetalias{millerZTFEarlyObservations2020}, the estimated correlation coefficient ($\rho\gtrsim0.5$ at a 40\% threshold) was likely overestimated due to aggregating the independently fitted parameters of individual SNe, as these two parameters are intrinsically degenerate \citepalias[][see also \citealp{gonzalez-gaitanRISETIMENORMAL2011}]{PaperI}. 
Across the entire population, the correlation between $t_\mathrm{rise}$ and the color evolution parameter $\alpha_g - \alpha_r$ remains weak. Finally, we recover the near-perfect correlation between $\alpha_g$ and $\alpha_r$ ($\rho\gtrsim0.95$) originally identified in \citetalias{millerZTFEarlyObservations2020}.

Under the hierarchical Bayesian framework, the posterior of an individual SN is determined by the likelihood as well as the population prior that is learned from the modeling process.
As expected, this approach significantly alleviates the spurious redshift-dependence of parameters induced by the Malmquist bias \citepalias{millerZTFEarlyObservations2020}, such that the inferred $t_\mathrm{rise}$ for individual events no longer correlates with the redshift ($\rho\simeq0$ and $p$-value $\simeq 0.6$). On the other hand, for SNe with poor data quality, the parameter posteriors largely reflect the population prior. Consequently, SNe in the distribution tails will be pulled back towards the population mean. Any statistical inference based on a specific subset of hierarchically modeled samples would require careful evaluation.

\subsection{The Robustness of the Hierarchical Bayesian Framework to Quality Cuts} \label{sec:quality_cuts}

\begin{figure}
    \centering
    \includegraphics[width=\linewidth]{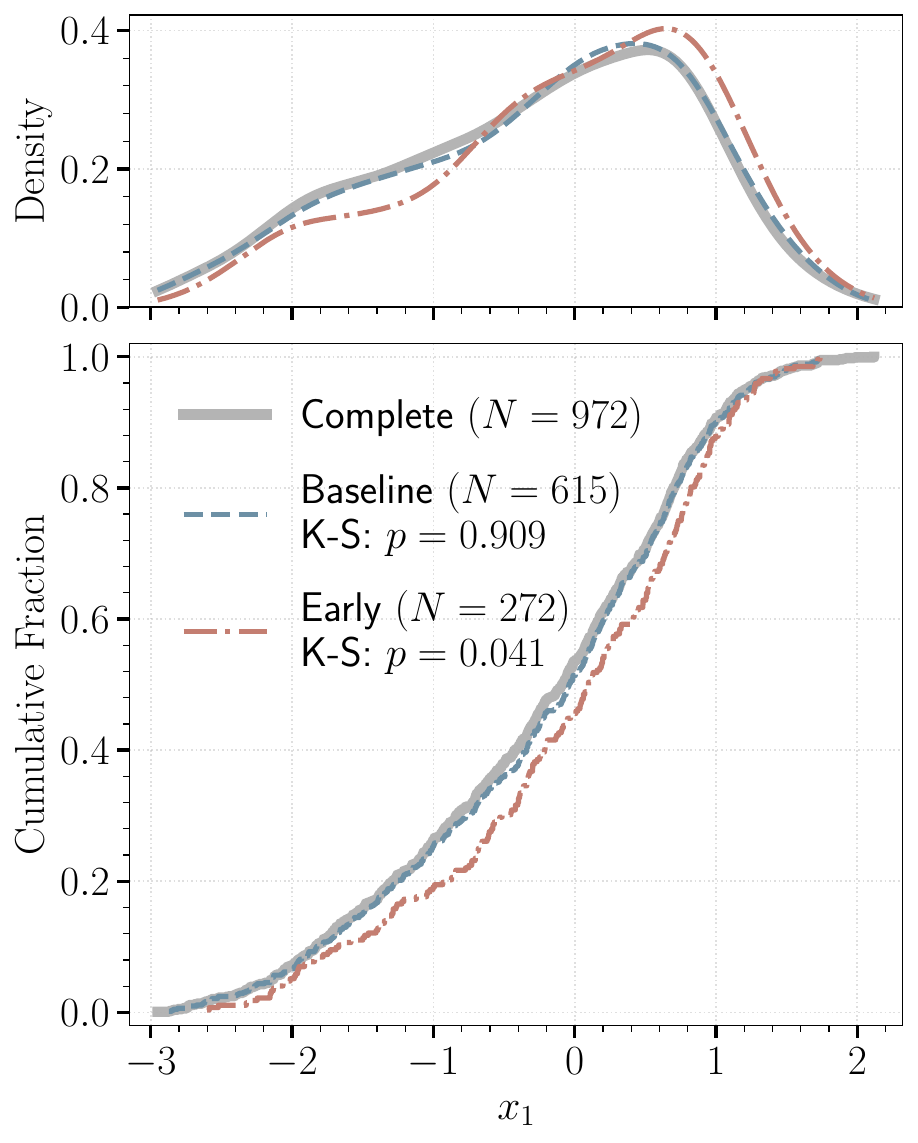}
    \caption{Additional early-time data-quality cuts may bias the sample towards higher-stretch SNe\,Ia. \textit{Top:} the kernel density estimation of the $x_1$ distribution for the volume-complete sample (black), the Baseline sample (blue), and the Early sample (orange). \textit{Bottom:} the cumulative distribution function of $x_1$ for the three samples. The K-S test $p$-values comparing the Baseline and Early samples against the volume-complete sample are listed.}
    \label{fig:ztf_quality_cuts_x1}
\end{figure}

Astronomical samples often rely on quality cuts to ensure reliable parameter inference. However, because the thresholds for these cuts are often arbitrary and vary significantly across studies, their systematic impact on population-level inferences is rarely quantified. The hierarchical framework reduces the need for strict data exclusions by naturally propagating measurement uncertainties and weighting individual events based on their constraining power \citepalias{PaperI}. This statistical coherency allows us to apply minimal quality cuts, thereby mitigating arbitrary selection biases while maintaining robust constraints on the underlying population properties.

To demonstrate this robustness, we apply our hierarchical framework to two subsets of the ZTF DR2 volume-complete sample (hereafter the Complete sample), which are defined by increasingly stringent quality cuts on early-time data coverage:
\begin{itemize}
    \item The Baseline sample (615): SNe with at least 10 pre-explosion non-detections (S/N $<5$) between $-100$\,days and $-25$\,days, ensuring sufficient data to constrain the baseline offset $C$.
    \item The Early sample (272): SNe in the Baseline sample with at least two nights of detections (S/N $>5$; multiple visits stacked over a single night) before $-10$\,days in both bands, ensuring reasonably good early-time coverage for more efficient sampling.
\end{itemize}

We note that even the Early sample is still relatively inclusive compared to the early-time coverage cuts commonly adopted in the literature. For instance, \citetalias{millerZTFEarlyObservations2020} only selected SNe\,Ia from the ZTF high-cadence fields with $\sim$nightly cadence, which typically yields $\ge$5 nights of detections before $-10$\,days for normal SNe\,Ia.

Applying additional quality cuts inevitably introduces bias. 
While the Baseline sample is statistically consistent with the Complete sample in the \texttt{SALT2} $x_1$ distribution, the Early sample is skewed towards higher $x_1$ values, with a Kolmogorov-Smirnov (K-S) test $p$-value of 0.041 (Figure~\ref{fig:ztf_quality_cuts_x1}) against the Complete sample. This is because higher-stretch SNe\,Ia have longer rise times \citep[e.g.,][]{ganeshalingamRisetimeDistributionNearby2011} and tend to be better sampled at early epochs ($<$$-10$\,days). This bias leads to an overestimation of the population mean of $t_\mathrm{rise}$ when limiting the inference to the Early sample alone.

\begin{deluxetable*}{lcccccc}
\tablecaption{Inferred Population Parameters for Early-Time ZTF SN\,Ia Light Curves across Varying Quality Cuts and Sub-samples \label{tab:quality_cuts}}
\tablewidth{0pt}
\tablehead{
\colhead{Parameter} & \colhead{Unit} & \colhead{Early} & \colhead{Baseline (early)} & \colhead{Complete (early)} & \colhead{Baseline} & \colhead{Complete}
}
\startdata
\multicolumn{7}{c}{\textbf{Population Means}} \\
$\mu_{{t_{{\rm rise}}}}$ & day & $18.85^{+0.12}_{-0.12}$ & $18.80^{+0.08}_{-0.08}$ & $18.75^{+0.07}_{-0.07}$ & $\mathbf{18.62}^{\mathbf{+0.10}}_{\mathbf{-0.10}}$ & $\underline{\mathbf{18.55}}^{\mathbf{+0.09}}_{\mathbf{-0.08}}$ \\
$\mu_{{\alpha_g}}$ & \nodata & $2.36^{+0.06}_{-0.06}$ & $2.33^{+0.04}_{-0.04}$ & $2.30^{+0.04}_{-0.04}$ & $2.33^{+0.05}_{-0.05}$ & $2.29^{+0.04}_{-0.04}$ \\
$\mu_{{\alpha_r}}$ & \nodata & $2.17^{+0.05}_{-0.05}$ & $2.13^{+0.04}_{-0.04}$ & $\mathbf{2.11}^{\mathbf{+0.03}}_{\mathbf{-0.03}}$ & $2.13^{+0.04}_{-0.04}$ & $\mathbf{2.10}^{\mathbf{+0.04}}_{\mathbf{-0.04}}$ \\
$\mu_{{\alpha_g-\alpha_r}}$ & \nodata & $0.19^{+0.02}_{-0.02}$ & $0.20^{+0.02}_{-0.02}$ & $0.20^{+0.02}_{-0.01}$ & $0.20^{+0.02}_{-0.02}$ & $0.20^{+0.02}_{-0.02}$ \\
[1ex]
\multicolumn{7}{c}{\textbf{Population Scatters}} \\
$\sigma_{{t_{{\rm rise}}}}$ & day & $1.45^{+0.10}_{-0.09}$ & $1.44^{+0.07}_{-0.06}$ & $1.42^{+0.06}_{-0.06}$ & $1.45^{+0.08}_{-0.07}$ & $1.42^{+0.07}_{-0.07}$ \\
$\sigma_{{\alpha_g}}$ & \nodata & $0.58^{+0.04}_{-0.04}$ & $0.59^{+0.03}_{-0.03}$ & $0.58^{+0.03}_{-0.03}$ & $0.59^{+0.04}_{-0.04}$ & $0.58^{+0.04}_{-0.04}$ \\
$\sigma_{{\alpha_r}}$ & \nodata & $0.49^{+0.04}_{-0.04}$ & $0.48^{+0.03}_{-0.03}$ & $0.48^{+0.03}_{-0.03}$ & $0.49^{+0.04}_{-0.03}$ & $0.48^{+0.03}_{-0.03}$ \\
$\sigma_{{\alpha_g-\alpha_r}}$ & \nodata & $0.17^{+0.02}_{-0.02}$ & $0.17^{+0.02}_{-0.02}$ & $0.16^{+0.02}_{-0.02}$ & $0.17^{+0.02}_{-0.02}$ & $0.17^{+0.02}_{-0.02}$ \\
[1ex]
\multicolumn{7}{c}{\textbf{Correlation Coefficients}} \\
$\rho(t_{{\rm rise}}, \alpha_g)$ & \nodata & $0.33^{+0.08}_{-0.08}$ & $0.30^{+0.06}_{-0.06}$ & $0.28^{+0.06}_{-0.06}$ & $\mathbf{0.18}^{\mathbf{+0.07}}_{\mathbf{-0.08}}$ & $\mathbf{0.16}^{\mathbf{+0.07}}_{\mathbf{-0.07}}$ \\
$\rho(t_{{\rm rise}}, \alpha_r)$ & \nodata & $0.36^{+0.08}_{-0.08}$ & $0.34^{+0.06}_{-0.06}$ & $0.33^{+0.06}_{-0.06}$ & $\mathbf{0.22}^{\mathbf{+0.07}}_{\mathbf{-0.08}}$ & $\mathbf{0.21}^{\mathbf{+0.07}}_{\mathbf{-0.07}}$ \\
$\rho(t_{{\rm rise}}, \alpha_g-\alpha_r)$ & \nodata & $0.09^{+0.12}_{-0.13}$ & $0.06^{+0.11}_{-0.10}$ & $0.03^{+0.10}_{-0.10}$ & $-0.03^{+0.12}_{-0.11}$ & $-0.06^{+0.11}_{-0.10}$ \\
$\rho(\alpha_g, \alpha_r)$ & \nodata & $0.97^{+0.01}_{-0.01}$ & $0.97^{+0.01}_{-0.01}$ & $0.97^{+0.01}_{-0.01}$ & $0.97^{+0.01}_{-0.01}$ & $0.97^{+0.01}_{-0.01}$ \\
\enddata
\tablecomments{Posterior medians and 16th--84th percentile intervals for the parameters using the $30\%$ threshold. The samples denote different data quality cuts. The Complete (early) and Baseline (early) columns correspond to the exact same events as the Early sample, but were modeled hierarchically alongside the lower-quality SNe in their respective broader sets. Bolded values indicate that the posterior difference between the given sample and the Early sample excludes zero at the $68\%$ credible level. Bolded and underlined values indicate exclusion at the $95\%$ credible level.}
\end{deluxetable*}

As a consistency check, we evaluate the identical subset of 272 SNe\,Ia that possess both baseline and early-time data coverage across all three sample configurations. Because these events contribute identical data to the likelihood in each model run, any variations in their inferred parameters isolate the influence of the population prior, which evolves as more sparsely sampled SNe are included. We find that the inferred population-level parameters for this subset remain statistically consistent (within the 68\% credible interval) across all three sample definitions (Table~\ref{tab:quality_cuts}), differing only by a marginally lower $\mu_{\alpha_r}$ when modeled within the Complete sample compared to fitting the Early sample alone.
Furthermore, the uncertainties on these parameters decrease when the subset is modeled alongside the full Baseline or Complete samples, compared to modeling the Early sample alone.
This demonstrates that the additional hundreds of sparsely sampled SNe\,Ia still contribute valuable constraining power to the overall population fit.

When comparing the full Baseline and Complete samples to the Early sample, population-level parameters remain broadly consistent. The only notable difference is a lower mean rise time ($-0.23$ and $-0.30$\,days for the Baseline and Complete samples, respectively), which weakens its correlation with rise indices. This decrease aligns with the Early sample's bias towards higher-stretch SNe\,Ia, underscoring the need to account for selection biases when applying quality cuts.

Furthermore, population properties of the Baseline and Complete samples agree at the 68\% credible level, demonstrating the robustness of our hierarchical framework: it effectively leverages limited data from sparsely sampled SNe, which independent unpooled models cannot fit, without biasing the overall inference.

\section{Correlations between Early-Time Observables and Full Light Curves} \label{sec:corr_x1}

As early-time light-curve properties primarily probe the outermost layers of the SN ejecta, the stretch of the full light curve (e.g., \texttt{SALT2} $x_1$) reflects the total ejecta mass, explosive energy, and \Ni yield. 
The \texttt{SALT2} $x_1$ distribution is bimodal \citep{nicolasRedshiftEvolutionUnderlying2021,larisonEnvironmentalDependenceType2024, ginolinZTFSNIa2025}, which could be attributed to two distinct populations of explosions. Searching for clustering in the phase space spanned by light-curve properties at different phases would provide independent evidence for the multimodality and shed light on a physical interpretation for the empirically derived \texttt{SALT2} parameters.
In this section, we systematically explore the correlation between early-time properties ($\alpha_r$, $\alpha_g - \alpha_r$, and $t_\mathrm{rise}$) and $x_1$.

We restrict the correlation analysis in this section to the subset of 272 SNe (modeled as part of the Complete sample) with good early-time and baseline coverage. This ensures robust constraints on individual early-time parameters, even though this subset is biased towards higher-stretch SNe\,Ia (Section~\ref{sec:quality_cuts}). While sparsely sampled SNe\,Ia contribute to population-level inferences, their individual posteriors are heavily influenced by the population prior. This strong prior dependence could artificially bias correlations with external parameters not explicitly included in the hierarchical model, such as $x_1$, due to Bayesian shrinkage.

\subsection{Rise Index and Early-Time Color Evolution}\label{sec:corr_x1_alpha}

\begin{figure*}[ht!]
    \centering
    \includegraphics[width=0.9\linewidth]{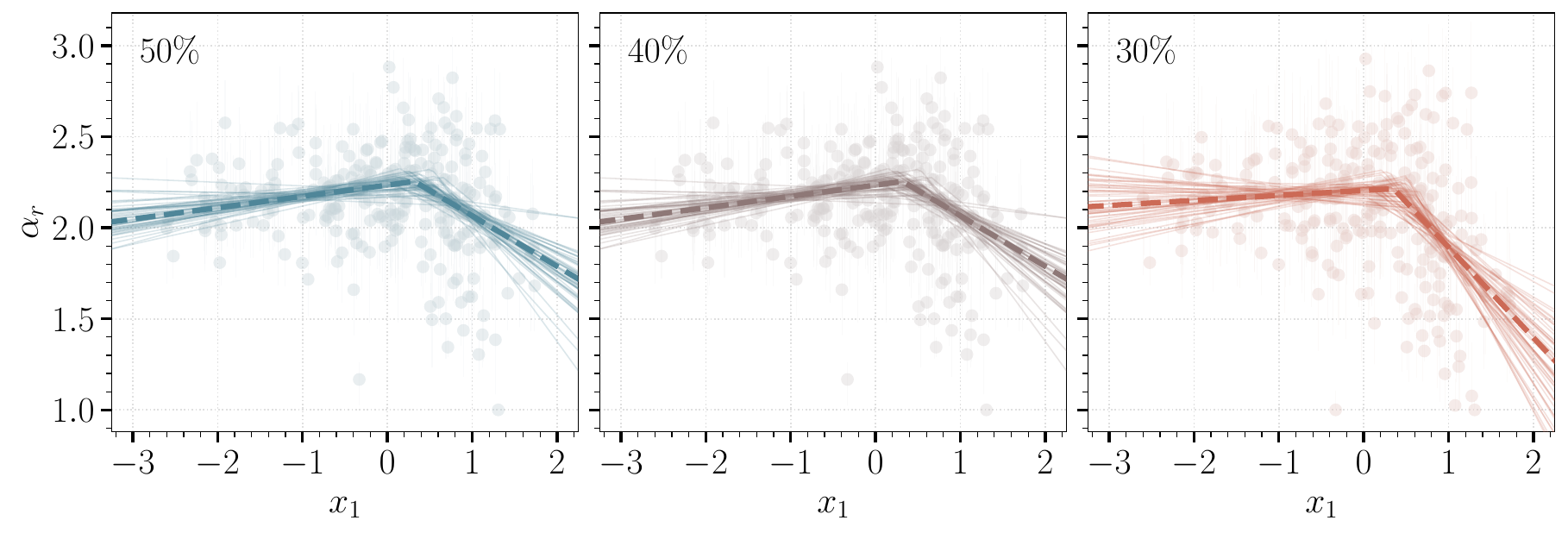}
    \caption{
        The inferred rise index ($\alpha_r$) exhibits a broken correlation with \texttt{SALT2} $x_1$, marked by a break at $x_\mathrm{brk} \gtrsim 0$. The panels show results from fitting early-time light curves truncated at 50\%, 40\%, and 30\% of peak flux. Dashed lines represent the median broken linear regression fits to the $\alpha_r$--$x_1$ relation. Fifty random posterior draws from the broken linear model are overplotted to illustrate uncertainties in the break point and slopes. For SNe\,Ia with $x_1 > x_\mathrm{brk}$, the rise index is negatively correlated with stretch, a trend that strengthens when employing a lower flux truncation threshold. Conversely, no significant correlation is observed for SNe with $x_1 < x_\mathrm{brk}$. Although the structural break is statistically robust, significant intrinsic scatter relative to the regression lines is observed across the entire $x_1$ range. 
    }
    \label{fig:ztf_alpha_x1_corr}
\end{figure*}

\begin{figure*}[ht!]
    \centering
    \includegraphics[width=0.9\linewidth]{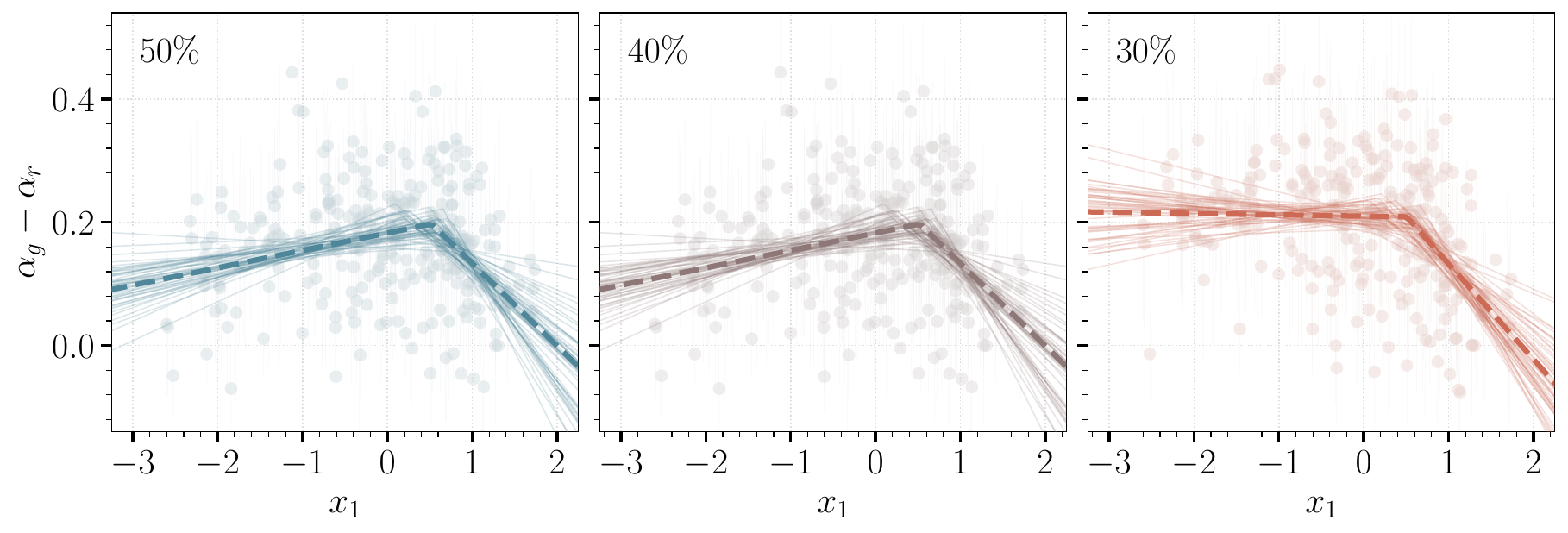}
    \caption{
        The early-time color evolution index ($\alpha_g - \alpha_r$) demonstrates a broken correlation with \texttt{SALT2} $x_1$, mirroring the behavior of the rise index seen in Figure~\ref{fig:ztf_alpha_x1_corr} despite the larger scatter. Posterior predictive formatting follows Figure~\ref{fig:ztf_alpha_x1_corr}. For light curves fit up to 30\% of peak flux, the inferred $\alpha_g - \alpha_r$ is negatively correlated with $x_1$ for SNe with $x_1 > x_\mathrm{brk}$, while no significant correlation is evident for $x_1 < x_\mathrm{brk}$. As higher truncation thresholds are applied, the negative correlation for the high-stretch population weakens and eventually vanishes, whereas the lack of correlation in the low-stretch population persists.
    }
    \label{fig:ztf_alpha_g-r_x1_corr}
\end{figure*}

Figures~\ref{fig:ztf_alpha_x1_corr} and \ref{fig:ztf_alpha_g-r_x1_corr} present the $\alpha_r$--$x_1$ and $(\alpha_g - \alpha_r)$--$x_1$ relations along with the posterior predictives from the broken linear models.
For the 30\% flux truncation threshold, both relations exhibit a clear structural break: among high-stretch SNe, $\alpha_r$ and $\alpha_g - \alpha_r$ are negatively correlated with $x_1$, implying that the highest-stretch SNe\,Ia tend to have near-linear rises that are much shallower than predicted by the fireball model, and show less pronounced early-time color evolution. In contrast, no significant trend is observed for the low-stretch population. 
As the truncation threshold increases, the anticorrelation for the high-stretch population weakens, suggesting that the strong correlation with stretch is fundamentally tied to the earliest epochs of emission. Conversely, the lack of correlation for the low-stretch population persists across all truncation thresholds. Despite the structural break, the continuous distribution of $\alpha_g - \alpha_r$ disfavors a strict division into distinct early-blue (persistently blue colors) and early-red (pronounced red$\rightarrow$blue color evolution) populations, as suggested in \citet{stritzingerRedBlueEarly2018} based on a smaller sample of well-observed SNe\,Ia. 

To locate the breaks in the $\alpha_r$--$x_1$ and $(\alpha_g - \alpha_r)$--$x_1$ relations and quantify their statistical significance, we implement a Bayesian errors-in-variables hierarchical regression model. 
To properly account for the measurement uncertainties for both parameters, we model latent ``true'' values for each individual SN\,Ia, indexed by $k$. The observed quantities are treated as samples drawn from Gaussian distributions centered on these latent values, such that 
\begin{equation}
    \begin{aligned}
    x_{1, k} &\sim \mathcal{N}(x_{1,\mathrm{true}, k}, \sigma_{x_1, k}^2) \\
    \psi_{k} &\sim \mathcal{N}(\psi_{\mathrm{true}, k}, \sigma_{\psi_{k}}^2)
    \end{aligned}
\end{equation}
where $\psi_{k}$ represents an early-time parameter for SN~$k$ (e.g., $t_{\mathrm{rise}}$, $\alpha_{r}$, $\alpha_{g} - \alpha_{r}$).
Here, $x_{1, k}$ and $\sigma_{x_1, k}$ come from the \texttt{SALT2} fits in the ZTF DR2 catalog, while the early-time parameter $\psi_{k}$ and its uncertainty $\sigma_{\psi_{k}}$ are derived from our hierarchical Bayesian modeling of the early-time light curves. 
Applying such a secondary hierarchical model to post-process posteriors derived from an initial hierarchical fit is a practical compromise, and future work should adopt a unified hierarchical model that simultaneously models the early-time light curves and their correlations with other properties, such as $x_1$.

The $(x_{1,\mathrm{true}, k}, \psi_{\mathrm{true}, k})$ pairs are then fit to a broken linear model with a piecewise linear relationship characterized by an initial slope ($\beta_1$) up to a break point ($x_\mathrm{brk}$), and a different slope ($\beta_2$) beyond that. This formalism is not intended to be a physical model, but rather an empirical, flexible tool to identify potential structural breaks in the correlation.
An intrinsic scatter $\epsilon$ is included in both models, which is assumed to be homoscedastic across the $x_1$ range. We assign uniform priors to the slopes and intercepts, weak half normal priors to the scatter, and a truncated normal prior to the break point $x_\mathrm{brk}\sim \mathcal{N}(0, 0.5^2)$, constrained to lie within $-1$ and 1.
To prevent individual SNe\,Ia with large uncertainties from dominating the inference of the break, we adopt a Student-T distribution likelihood for the regression, which is more robust to outliers. The degree of freedom $\nu$ is a free parameter to be inferred from the data, with a Gamma-distribution prior of $\Gamma(2, 0.1)$. 
We perform inference using the NUTS sampler, running 4 parallel chains with 2,000 draws each.

\begin{deluxetable}{cccc}
    \tablecaption{Posterior Inference for the Broken Linear Model in the Early-Time Properties (30\% Truncation Threshold) -- $x_1$ Relations.
    \label{tab:broken_params}}
    \tablehead{
    \colhead{Parameter} & \colhead{$t_\mathrm{rise}$\,[day]} & \colhead{$\alpha_r$} & \colhead{$\alpha_g - \alpha_r$}
    }
    \startdata
    $\beta_0$ & $19.10^{+0.11}_{-0.10}$ & $2.21^{+0.04}_{-0.04}$ & $0.21^{+0.01}_{-0.01}$ \\
$\beta_1$ & $1.20^{+0.10}_{-0.09}$ & $0.03^{+0.04}_{-0.04}$ & $-0.00^{+0.01}_{-0.01}$ \\
$\beta_2$ & $0.17^{+0.20}_{-0.26}$ & $-0.50^{+0.11}_{-0.14}$ & $-0.15^{+0.04}_{-0.05}$ \\
$\beta_1 - \beta_2$ & $1.05^{+0.27}_{-0.25}$ & $0.53^{+0.15}_{-0.12}$ & $0.15^{+0.05}_{-0.04}$ \\
$x_b$ & $0.20^{+0.19}_{-0.13}$ & $0.35^{+0.17}_{-0.18}$ & $0.51^{+0.17}_{-0.23}$ \\
$\epsilon$ & $0.31^{+0.08}_{-0.07}$ & $0.18^{+0.04}_{-0.04}$ & $0.04^{+0.01}_{-0.01}$ \\
$\nu$ & $2.6^{+1.0}_{-0.7}$ & $5.0^{+4.9}_{-2.0}$ & $18.2^{+16.2}_{-9.5}$ \\
    \enddata
    \end{deluxetable}

The broken linear model identifies a stark transition in both relations: the initial flat slope ($\beta_1\simeq0$) transitions to a significantly negative slope ($\beta_2 < 0$) beyond a break point for both the $\alpha_r$--$x_1$ and $(\alpha_g - \alpha_r)$--$x_1$ relations. The location of the break point is consistent across both relations.
Draws from the posterior predictives of the broken linear model are overplotted in Figure~\ref{fig:ztf_alpha_x1_corr}. The fitting results for the broken linear model are summarized in Table~\ref{tab:broken_params}.

Despite the structural break, the intrinsic scatter ($\epsilon\simeq0.18$ for $\alpha_r$ and $\epsilon\simeq0.04$ for $\alpha_g - \alpha_r$) dominates the uncertainty budget for individual SNe\,Ia. Therefore, the anticorrelation with $x_1$ in the high-stretch population should be interpreted as a population-level trend rather than a deterministic relationship for individual SNe\,Ia. The degrees of freedom ($\nu$) for the Student-T likelihoods are poorly constrained for both models, primarily due to the large relative uncertainties in the rise-index measurements. 

Our findings provide physical context for the connection between early-time rise morphology, color evolution, and light-curve stretch. Previous studies have noted that, within a week of first light, the brightest, slowest-evolving SNe\,Ia exhibit flatter early color evolution \citep[remaining persistently blue or showing a ``blue plateau'';][]{wuCommonOriginNormal2025}, whereas faster-evolving events undergo a pronounced red$\rightarrow$blue transition \citep{stritzingerRedBlueEarly2018, bullaZTFEarlyObservations2020}. Our analysis quantifies this relationship, demonstrating that persistent blue colors are directly coupled with lower initial rise indices in the first few days post-explosion. By utilizing a volume-complete sample, we further reveal that these color--stretch trends are primarily driven by the high-stretch population, which is often overrepresented in flux-limited surveys, while the low-stretch population does not follow similar trends.

\subsection{Rise Time}\label{sec:corr_x1_t_rise}

\begin{figure*}[ht!]
    \centering
    \includegraphics[width=0.9\linewidth]{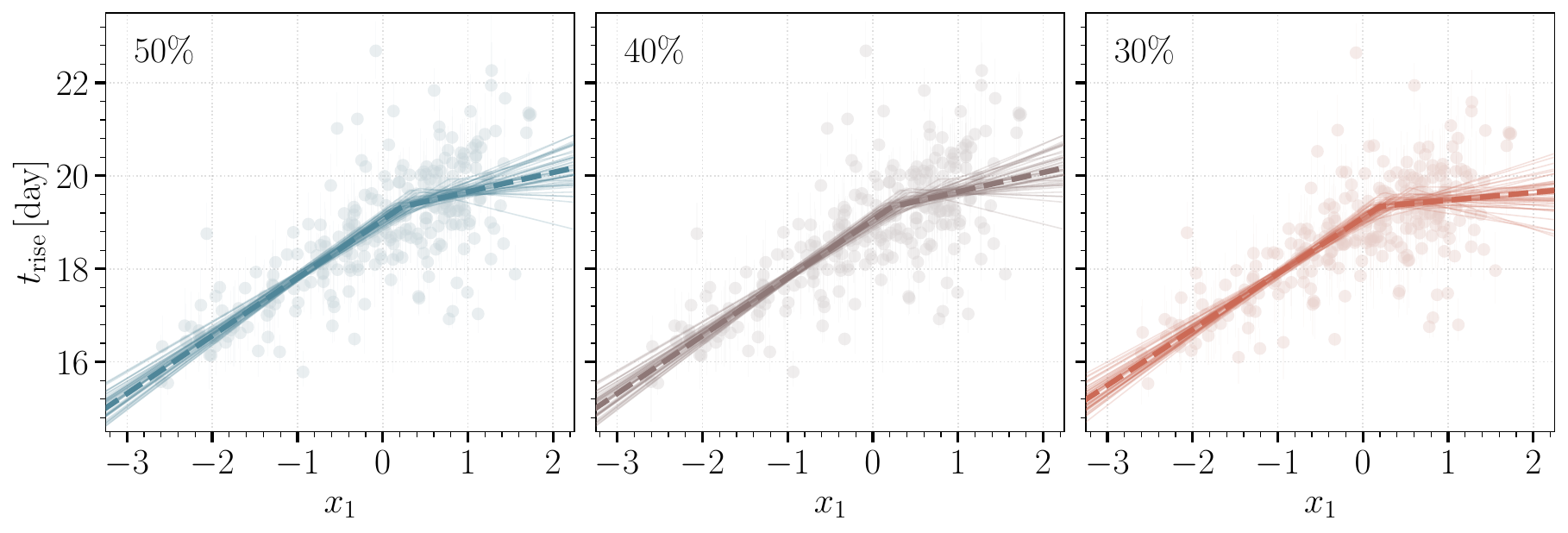}
    \caption{
        The rise time, $t_\mathrm{rise}$, exhibits a broken positive correlation with the full light-curve stretch (\texttt{SALT2} $x_1$) across ZTF SNe\,Ia. A break occurs near $x_\mathrm{brk} \gtrsim 0$: for SNe\,Ia with $x_1 > x_\mathrm{brk}$, the rise time increases more slowly with $x_1$ than for those with $x_1 < x_\mathrm{brk}$. Posterior predictive formatting follows Figure~\ref{fig:ztf_alpha_x1_corr}.
    }
    \label{fig:ztf_t_rise_x1_corr}
\end{figure*}

\begin{figure}[ht!]
    \centering
    \includegraphics[width=0.97\linewidth]{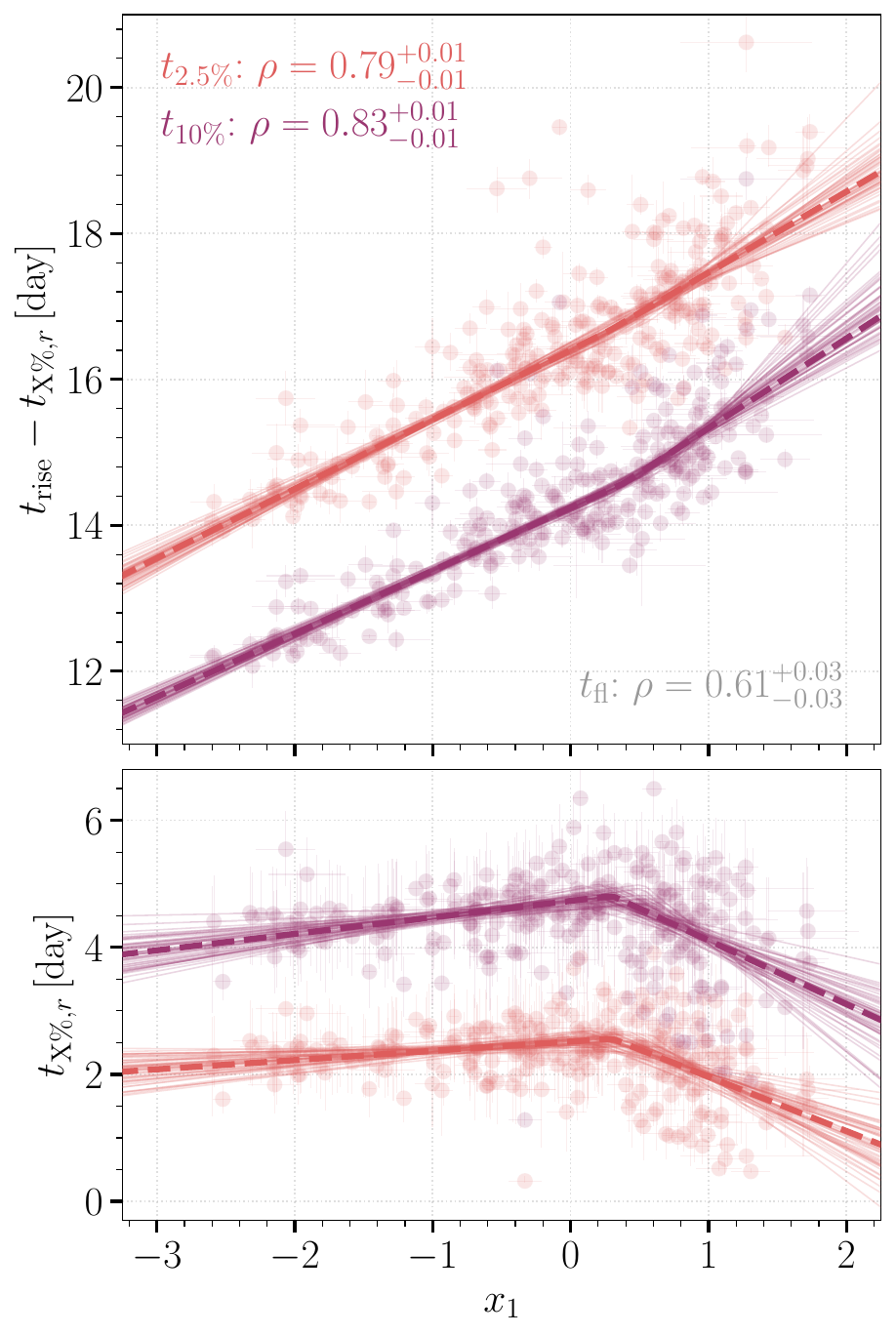}
    \caption{
        The adjusted rise time reveals a continuous, strongly correlated relationship with the full light-curve stretch $x_1$, eliminating the break seen in the standard $t_\mathrm{rise}$ measurement. 
        \textit{Top}: The adjusted rise time, measured from the 2.5\% and 10\% peak flux epochs in the $r$-band, exhibits a tighter correlation with $x_1$ than the traditional rise time from first light ($t_\mathrm{rise}$). This is seen in the substantial increase in the Pearson correlation coefficient $\rho$. We overlay the posterior predictive distributions of a broken linear regression model, applying a strong prior on $x_\mathrm{brk}$ based on the original $t_\mathrm{rise}$--$x_1$ relation; no significant break is recovered for the adjusted relation.
        \textit{Bottom}: The duration from the extrapolated first light to these early fractional flux epochs ($t_{\mathrm{X}\%, r}$) displays a broken correlation with $x_1$, turning over near $x_1\simeq 0$. The overplotted posterior predictives of a broken linear regression model recover breaking points consistent with the original $t_\mathrm{rise}$--$x_1$ trend. %This distinct early-time behavior phenomenologically drives the break observed in the total rise time relation.
    }
    \label{fig:ztf_t_rise_x1_corr_adjusted}
\end{figure}

Figure~\ref{fig:ztf_t_rise_x1_corr} displays a strong positive correlation between $t_\mathrm{rise}$ and $x_1$ for ZTF SNe\,Ia in the volume-complete sample, consistent with previous studies \citep{riessRiseTimeNearby1999,goldhaberTimescaleStretchParameterization2001,conleyRiseTimeType2006,haydenRISEFALLTYPE2010,ganeshalingamRisetimeDistributionNearby2011,gonzalez-gaitanRISETIMENORMAL2011}. Notably, the $t_\mathrm{rise}$--$x_1$ relation also exhibits a break: for SNe\,Ia with $x_1>x_\mathrm{brk}$, the average rise time increases more slowly with $x_1$ than for those with $x_1 < x_\mathrm{brk}$.

Applying the same Bayesian regression analysis described in Section~\ref{sec:corr_x1_alpha}, we confirm a significant decrease in slope across the break point at $x_\mathrm{brk} \simeq 0.2$ in the $t_\mathrm{rise}$--$x_1$ relation, broadly consistent with the break points identified in the $\alpha_r$--$x_1$ and $(\alpha_g - \alpha_r)$--$x_1$ relations. This structural break is most prominent for the 30\% truncation threshold. Despite the clear slope transition, intrinsic scatter ($\epsilon \simeq 0.3$\,days) still dominates the uncertainty for individual SNe\,Ia across the entire $x_1$ range. Furthermore, the Student-T likelihood strongly favors a low degree of freedom ($\nu \simeq 2.6$), indicating that significant outliers drive heavy-tailed residuals with respect to the broken linear model. The fitting results are summarized in Table~\ref{tab:broken_params}. 

Part of this behavior, however, could arise from the light-curve fitting process. In fact, both statistical uncertainties and systematic errors in the inferred $t_\mathrm{rise}$ are dominated by the extrapolation under the assumption of a power-law rise. 
Statistically, the degeneracy between $t_\mathrm{rise}$ and $\alpha$ in the power-law model elongates the posterior distribution, which inflates the uncertainty in the inferred $t_\mathrm{fl}$, and thus $t_\mathrm{rise}$. This large uncertainty in $t_\mathrm{rise}$ potentially masks its intrinsic correlation with $x_1$.
Physically, for SNe with a significant dark phase \citep[e.g., due to a highly confined \Ni distribution;][]{piroWHATCANWE2013,piroCONSTRAINTSSHALLOW562014}, the $t_\mathrm{fl}$ inferred from a power-law fit may poorly represent the true explosion time. 

To mitigate both the statistical uncertainties of extrapolation and these physical ambiguities, we introduce an adjusted rise time, defined as the duration between the epoch when the flux reaches a specific fraction (X\%) of its peak and the epoch of maximum light. This metric is more robustly constrained by the data and less model-dependent than the standard rise time (which implicitly assumes X$\%=0$). We denote the interval from $t_\mathrm{fl}$ to the X\% peak flux epoch in filter $b$ as $t_{\mathrm{X}\%, b}$. Under our assumed power-law rise, this interval is:
\begin{equation}\label{eq:t_x_percent}
    t_{\mathrm{X}\%, b} = t_\mathrm{pivot}\left(\frac{\mathrm{X}}{A_b}\right)^{1/\alpha_b},
\end{equation}
making the adjusted rise time $t_\mathrm{rise} - t_{\mathrm{X}\%, b}$. Discarding the earliest phases of the light curve allows this metric to better isolate the diffusion timescale of the bulk ejecta, reflecting the total ejecta mass and \Ni yield, while $t_{\mathrm{X}\%, b}$ remains sensitive to the \Ni distribution near the surface. Because $x_1$ in the ZTF DR2 sample is derived using only data within $[-10, +40]$ rest-frame days from peak \citep{rigaultZTFSNIa2025a}, and the \texttt{SALT2} model often fails to capture the full diversity of light curves before $-10$\,days \citep{amenoucheZTFSNIa2025}, we naturally expect $t_\mathrm{rise} - t_{\mathrm{X}\%, r}$ to exhibit a stronger correlation with $x_1$ than $t_\mathrm{rise}$.

Figure~\ref{fig:ztf_t_rise_x1_corr_adjusted} confirms this expectation, showing the relations between $x_1$ and both $t_\mathrm{rise} - t_{\mathrm{X}\%, r}$ and $t_{\mathrm{X}\%, r}$ for $\mathrm{X}\%=2.5\%$ and $10\%$, with a fixed truncation threshold at $30\%$ of the peak flux. The adjusted rise time indeed correlates more strongly with $x_1$ than the standard $t_\mathrm{rise}$, yielding a substantial increase in the Pearson correlation coefficient ($\rho$ rises from $0.6$ to $0.8$).

With the same Bayesian regression analysis, we can no longer identify a significant break within $x_1 \in [-1, 1]$. By imposing a strong prior derived from the posterior of the original $t_\mathrm{rise}$--$x_1$ regression on the break location ($x_\mathrm{brk} \sim \mathcal{N}(0.20, 0.16)$), we determine that for $\mathrm{X}\%=2.5\%$, the difference in slopes across the break is completely consistent with zero ($\beta_1 - \beta_2 = -0.16 \pm 0.19$). This confirms the absence of a structural break in the relation between the adjusted rise time and $x_1$. However, the substantial intrinsic scatter ($\epsilon \simeq 0.3$\,days) still dominates the uncertainty for individual SNe\,Ia, indicating that even the adjusted rise time is not perfectly coupled to the overall light-curve stretch. Consequently, predicting the rise time based solely on $x_1$ is subject to considerable systematic uncertainty. Finally, for $\mathrm{X}\%=10\%$, similar analysis yields marginal but statistically weak evidence ($\beta_1 - \beta_2 = -0.37 \pm 0.22$) for a reversed break, where the relation actually steepens beyond $x_\mathrm{brk} \simeq 0.2$.

The bottom panel of Figure~\ref{fig:ztf_t_rise_x1_corr_adjusted} demonstrates that the break in the original $t_\mathrm{rise}$--$x_1$ relation is primarily driven by the behavior of $t_{\mathrm{X}\%, r}$. Within the high-stretch population, $t_{\mathrm{X}\%, r}$ and $x_1$ are anticorrelated: as stretch increases within this regime, SNe\,Ia generally reach the early flux thresholds faster. Conversely, for the low-stretch population, $t_{\mathrm{X}\%, r}$ remains nearly constant across a wide range of $x_1$, with a mild positive correlation. 

Previous studies have suspected differences in the rise times between high- and low-stretch SNe\,Ia. For instance, analysis of the Lick Observatory Supernova Search sample showed that slow-declining SNe\,Ia rise faster than expected from a standard single-parameter stretch family \citep{ganeshalingamRisetimeDistributionNearby2011}. Similar hints were found when comparing the aggregate stretch-corrected rise times of high- and low-stretch SNe\,Ia in the Supernova Legacy Survey sample \citep{gonzalez-gaitanRISETIMENORMAL2011}, though not statistically significant. Our population-level measurement confirms the dichotomy in the rise time--stretch relation, which demonstrates that a single stretch parameter does not capture the full diversity of early SN\,Ia light curves \citep{haydenRISEFALLTYPE2010,gonzalez-gaitanRISETIMENORMAL2011}.

\subsection{Tests for Population Prior Effects}

A primary caveat of the analysis above is that the lowest-stretch SNe\,Ia have systematically fewer early detections before reaching the truncation threshold than those with broader light curves and longer rise times. As a result, the hierarchical structure and inferred population distribution exert a stronger influence on their early-time parameter posteriors compared to the well-sampled, high-stretch events. For instance, the $\alpha$ posteriors for these fast-declining SNe are pulled toward the population mean due to Bayesian shrinkage, which might mask a genuine anticorrelation between $x_1$ and $\alpha$ at the low-stretch end, should lower-stretch SNe\,Ia intrinsically have steeper rise indices.

A definitive solution would require more sophisticated population priors (e.g., Gaussian mixture or non-parametric models) to explicitly capture the multimodality in the early-time parameter distributions and the correlation structure within each sub-population. However, this approach demands a substantially larger sample size to constrain the additional hyperparameters and is beyond the scope of this work. Instead, we perform a straightforward test by applying the hierarchical Bayesian framework separately to the two sub-populations, divided at $x_\mathrm{brk}=0.2$, to independently infer their population-level properties. If Bayesian shrinkage is masking underlying trends, modeling the low-$x_1$ population separately should reveal a negative correlation between $\alpha$ and $x_1$, as the individual posteriors would no longer be pulled toward a single global population mean.

For the $x_1>0.2$ population, we consistently recover significant negative correlations ($\rho \simeq -0.5$) between $x_1$ and the early-time parameters $t_{2.5\%, r}$, $\alpha_r$, and $\alpha_g - \alpha_r$. In contrast, the $x_1<0.2$ population shows moderate positive correlations between $x_1$ and both $\alpha_r$ and $\alpha_g - \alpha_r$ ($\rho \simeq 0.3$), as well as a strong positive correlation between $x_1$ and $t_{2.5\%, r}$ ($\rho \simeq 0.7$) that is weakly detected in Figure~\ref{fig:ztf_t_rise_x1_corr_adjusted}. These findings confirm that the behavior of early-time observables relative to the full light-curve stretch is fundamentally distinct between low- and high-stretch populations.

Alternatively, \citetalias{PaperI} proposed fitting individual SN light curves using a custom prior derived from the full hierarchical population prior. This tailored prior preserves the correlation structure of the key early-time parameters but isolates the mean and scatter of the nuisance parameter $\ln A$, minimizing the over-regularization of $t_\mathrm{rise}$ and $\alpha$. In Appendix~\ref{sec:ztf_pop_prior}, we apply this custom population prior to the subset of 272 well-sampled SNe\,Ia. We find that the structural breaks in the correlations between $x_1$ and the early-time parameters remain consistent with those derived from the primary hierarchical model, although the measurement uncertainties for individual SNe\,Ia are generally larger. This consistency suggests that the observed correlations are not artifacts of the population prior, but rather reflect genuine physical trends in the data.

\section{Discussion}\label{sec:discussion}

\subsection{Spectroscopically Peculiar SNe\,Ia}\label{sec:discussion_peculiar}

While our analysis has primarily focused on spectroscopically normal SNe\,Ia (including 91T-like events), our hierarchical Bayesian framework can be readily applied to other spectroscopic subclasses. In this section, we briefly explore three common subtypes of spectroscopically peculiar SNe\,Ia in the ZTF DR2 sample: 91bg-like \citep{filippenkoSubluminousSpectroscopicallyPeculiar1992}, 02cx-like \citep{liSN2002cxMost2003}, and 03fg-like events \citep{howellTypeIaSupernova2006}. Rarer subclasses are excluded due to their limited sample sizes ($<$10) in ZTF DR2 \citep{dimitriadisZTFSNIa2025,alburaiZTFSNIa2026}.

Given the substantial internal diversity and smaller sample sizes of peculiar SNe\,Ia, defining a volume-complete sample for each subclass is nontrivial. For the intrinsically bright, slow-evolving 03fg-like SNe, we assume they are volume-complete within the same $z < 0.06$ threshold used for the normal sample, yielding 11 events. Conversely, the fainter, fast-evolving 91bg-like and 02cx-like SNe are underrepresented at this distance. To construct a more representative sample for these subclasses, we apply a stricter redshift cut of $z < 0.038$, which corresponds to a limiting absolute magnitude that is $1$\,mag fainter. This cut secures 41 91bg-like and 12 02cx-like SNe\,Ia for our analysis, though we acknowledge that the faint and fast extremes of both subclasses are still underrepresented, and the inferred rise times are likely biased towards longer values.

\begin{figure}
    \centering
    \includegraphics[width=\linewidth]{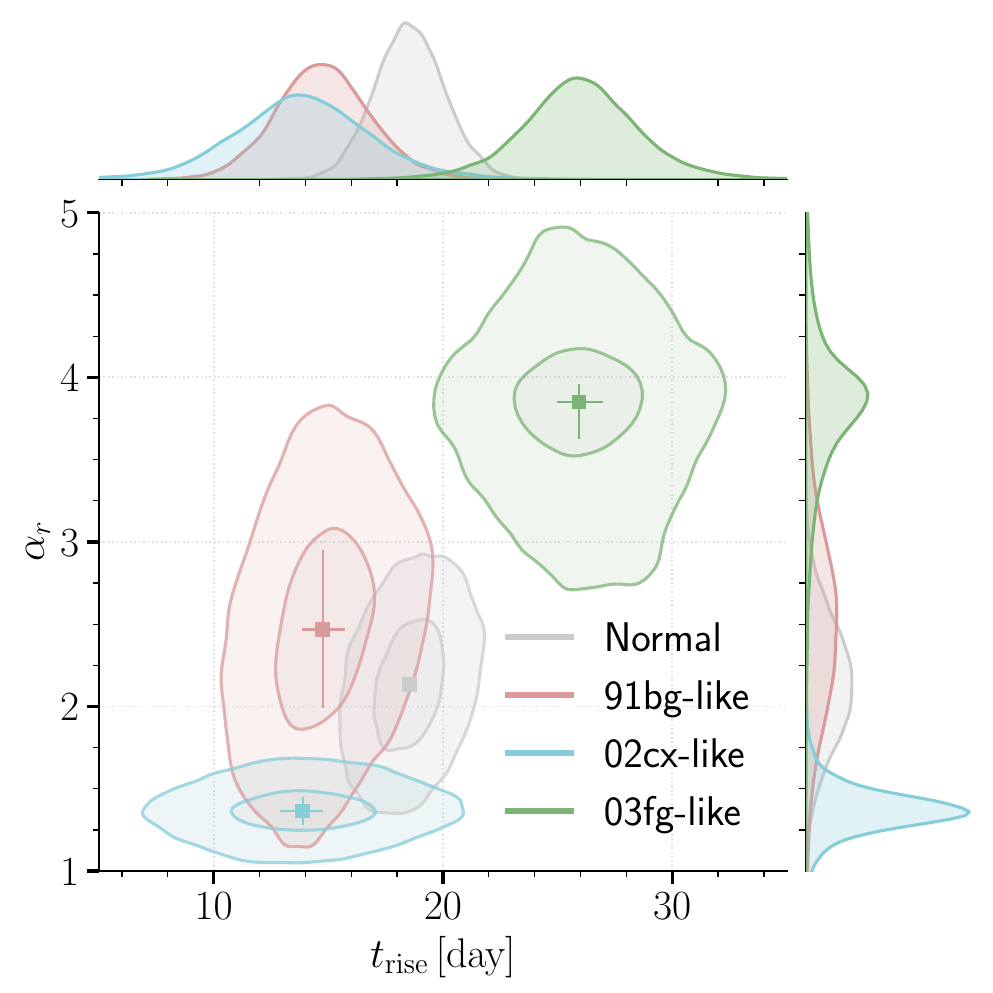}
    \caption{
        Early-time properties of spectroscopically peculiar SNe\,Ia in the ZTF DR2 sample compared to the normal population (gray). Subluminous and fast-evolving 91bg-like (red) and 02cx-like (cyan) events generally exhibit shorter adjusted rise times than normal SNe\,Ia, whereas overluminous and slow-evolving 03fg-like events (green) display longer rise times. Furthermore, 02cx-like and 03fg-like events present notably low and high rise indices ($\alpha_r$), respectively. 
        The contours represent the posterior predictive distributions for each subclass, showing the 39.3\% and 86.5\% credible regions (equivalent to 1$\sigma$ and 2$\sigma$ in a 2D Gaussian). Overlaid are the medians and the 16th--84th percentile ranges (points and errorbars) of the inferred population means.
    }
    \label{fig:ztf_peculiar}
\end{figure}

For each subclass, we apply our hierarchical Bayesian framework to infer the population-level properties of $t_\mathrm{rise}$ and $\alpha_r$, utilizing a 40\% peak flux truncation threshold to include more early-time data for robustness. In Figure~\ref{fig:ztf_peculiar}, we compare these inferred population-level properties against those of normal SNe\,Ia. The contours map the posterior predictive distributions for each subclass, with the medians and 16th--84th percentile intervals of the inferred population means overplotted as the points and errorbars, respectively. 

The 03fg-like and 02cx-like events clearly separate from the normal population in the $t_\mathrm{rise}$--$\alpha_r$ plane. The 03fg-like events display extended rise times ($\mu_{t_\mathrm{rise}}\simeq26$\,days) and much steeper rise indices ($\mu_{\alpha_r}\simeq3.9$), whereas the latter exhibit significantly shorter rise times ($\mu_{t_\mathrm{rise}}\simeq14$\,days) and shallower rise indices ($\mu_{\alpha_r}\simeq1.4$). As demonstrated in \citetalias{PaperI}, modeling an early flux excess with a single power-law rise can artificially bias the inferred early-time parameters: a short-duration ($\lesssim$2\,days) excess biases $t_\mathrm{rise}$ longer and $\alpha$ higher, while a long-duration ($\sim$5\,days) excess biases $t_\mathrm{rise}$ shorter and $\alpha$ lower. Consequently, the distinct parameters of 03fg-like and 02cx-like events could be phenomenologically attributed to the presence of short- and long-duration early flux excesses, respectively.

Spiky early flux excesses are commonly seen among 03fg-like events \citep[e.g.,][]{jiangDiscoveryFastestEarly2021,dimitriadisSN2021znyEarly2023,srivastavLuminousTypeIa2023,siebertGroundbasedJWSTObservations2023}. The leading explosion scenario involves a violent double WD merger \citep[][cf. \citealp{taubenbergerSN2012dnEarly2019,dimitriadisSN2021znyEarly2023,kwokGroundbasedJWSTObservations2024} for independent observational evidence]{pakmorNORMALTYPEIa2012}. The tidally disrupted secondary WD leaves dense, confined CSM in the vicinity of the SN \citep{raskinTIDALTAILEJECTION2013,raskinTYPEIaSUPERNOVAE2014}, and the consequent ejecta--CSM interaction powers a prominent brief early excess \citep{piroUSINGDOUBLEPEAKEDSUPERNOVA2015,noebauerTypeIaSupernovae2016,piroEXPLORINGPOTENTIALDIVERSITY2016}.  Our findings provide additional statistical support for this scenario and confirm that short-duration early flux excesses are likely prevalent among 03fg-like events.

The short rise times and shallow rise indices of 02cx-like events have been identified in previous studies \citep[e.g.,][]{mageeTypeIaxSupernova2016,barnaSN2019mujWellobserved2021,maguireSN2020udySN2023,hoogendamEarlyExtensiveUltraviolet2025,mageeSN2024bfuSN2025,singhPhotometrySpectroscopySN2026}. These properties could be well reproduced by a partial deflagration of a near-$\Mch$ WD \citep{kromer3DDeflagrationSimulations2013,finkThreedimensionalPureDeflagration2014}, in which the \Ni is well mixed into the outer ejecta layers, leading to a shallow rise index of $\alpha\lesssim1.4$ \citep{noebauerEarlyLightCurves2017}. The low ejecta mass following a weak deflagration, on the other hand, results in a short diffusion timescale and thus a short rise time.

For the 91bg-like events, the mean $\alpha_r$ remains poorly constrained despite a larger sample size, driving a broad spread in its posterior predictive distribution. Nevertheless, their rise times are consistently shorter than those of normal SNe\,Ia ($\mu_{t_\mathrm{rise}}\simeq15$\,days), which aligns with findings from previous studies \citep{zhengEmpiricalFittingMethod2017a}.

\subsection{Anomalous Early Emission Signatures}\label{sec:discussion_anomalies}
Having identified distinct phase-space locations for the spectroscopically peculiar subclasses and connected their distinctive early-time properties to the presence of early flux excesses, we now turn to the normal SNe\,Ia. In this section, we analyze outliers in the relations discussed in Section~\ref{sec:corr_x1} to search for signatures of early flux excesses and other anomalous behaviors within the ZTF DR2 volume-complete sample. The analysis is again limited to the Early subset from the Complete sample, which should be more sensitive to early-time anomalies.

\subsubsection{Steep/Long Risers: Signatures of Companion/CSM Interaction}\label{sec:discussion_anomalies_long}

\begin{figure*}[ht!]
    \centering
    \includegraphics[width=\linewidth]{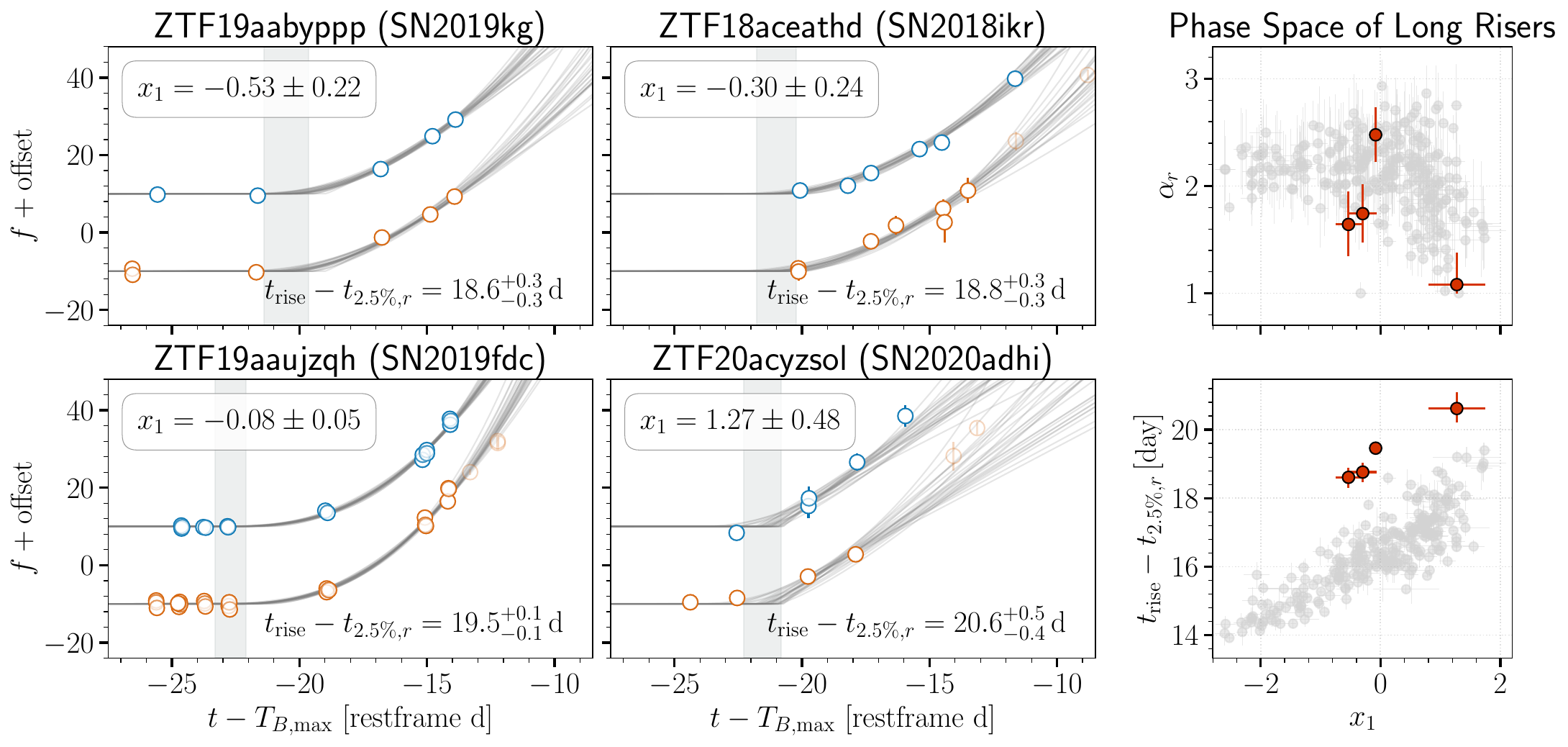}
    \caption{
        Light curve fitting results for SNe\,Ia consistent with an anomalously large $t_\mathrm{rise} - t_\mathrm{2.5\%, r}$ compared to other SNe with similar $x_1$ in ZTF DR2 and their location in the phase diagrams. \textit{Left}: The blue and orange points represent the observed fluxes in ZTF $g$ and $r$ bands, respectively. The solid curves are the posterior predictive light curves from the hierarchical Bayesian model. The shaded regions indicate the 16th--84th percentile range of the inferred $t_\mathrm{fl}$. The SNe are ordered by their $x_1$ values.
        \textit{Right}: The location of these four events in the $\alpha_r$--$x_1$ plane and the $t_\mathrm{rise} - t_{2.5\%, r}$--$x_1$ plane, with the rest of the sample shown in gray for comparison.
    }
    \label{fig:ztf_anomalies_long}
\end{figure*}

As discussed in Section~\ref{sec:discussion_peculiar}, ejecta--CSM/companion collisions can produce short-lived ($\lesssim$2\,days) early flux excesses, which manifest as steep rises and/or extended rise times when fitted with a power-law rise \citepalias{PaperI}.

Within our volume-complete sample, however, we find no SNe with $\alpha_r$ values exceeding the median of their respective $x_1$ bins by 2$\sigma$ (i.e., $\alpha_r\gtrsim3.2$ for $x_1>0$ and $\alpha_r\gtrsim3.0$ for $x_1<0$). However, the population prior applied to the individual $\alpha$ posteriors may artificially suppress extreme outliers. To test this, we refit the light curves of the 03fg-like events within $z=0.06$ (Section~\ref{sec:discussion_peculiar}), forcing the population priors to match those of the normal SNe\,Ia. Despite presenting a population mean of $\mu_{\alpha_r} \simeq 3.9$ when these events are modeled independently, only one 03fg-like event (SN\,2018gdl) yields an $\alpha_r>3$ under the normal priors. Even when expanding the sample to include all DR2 03fg-like events with good baseline coverage and $N_\mathrm{early}\ge2$, SN\,2018gdl remains the sole event confidently exhibiting $\alpha_r>3$. This confirms that Bayesian shrinkage disproportionately affects high-$\alpha$ SNe under typical ZTF cadences \citepalias{PaperI}, making the inferred $\alpha$ an unreliable metric for identifying early flux excesses in our sample.

In contrast, the rise time is less sensitive to the Bayesian shrinkage effect. Among the eleven 03fg-like events that pass our baseline and early detection cuts, eight have an adjusted rise time of $t_\mathrm{rise} - t_{2.5\%, r} > 20$\,days, clearly separating them from the normal population (Figure~\ref{fig:ztf_t_rise_x1_corr_adjusted}). Among the normal SN\,Ia sample, we identify four SNe\,Ia with $t_\mathrm{rise} - t_{2.5\%, r}$ values that are more than 3$\sigma$ above the median of their respective $x_1$ bins.
These events, which include SN\,2018ikr, SN\,2019fdc, SN\,2019kg, and SN\,2020adhi, are highlighted in Figure~\ref{fig:ztf_anomalies_long}.\footnote{If the same criterion is applied to the standard $t_\mathrm{rise}$ metric, only SN\,2019fdc remains a significant outlier ($t_\mathrm{rise}=22.6$\,days).} 
Loosening the threshold to 2.5$\sigma$ reveals an additional event, SN\,2020skp.
% For all but SN\,2020skp, the \texttt{SALT2} model systematically underestimates the flux prior to $-10$\,days relative to peak.\footnote{\texttt{SALT2} fits can be found at \url{https://ztfcosmo.in2p3.fr/}.} 

SN\,2018ikr, SN\,2019kg, SN\,2019fdc, and SN\,2020skp all exhibit extended adjusted rise times of $\sim$19\,days and feature weak early-time color evolution ($\alpha_g - \alpha_r\simeq0.0$--0.1). Because their $x_1$ values are typical and their light curves align well with the \texttt{SALT2} model near peak,\footnote{\texttt{SALT2} fits can be found at \url{https://ztfcosmo.in2p3.fr/}.} these anomalously long rises do not stem from a fundamental shift in overall light-curve morphology. Instead, they point to distinct early-time behavior likely decoupled from later evolutionary phases. In fact, for all but SN\,2020skp, the \texttt{SALT2} model systematically underestimates the flux prior to $-10$\,days relative to peak.
Although their $\alpha_r$ values span a broad range ($\sim$1.5--2.5), a short-duration ($\lesssim$2\,days) early flux excess, subject to varying sampling and noise, can readily reproduce this spread in $\alpha$ while consistently biasing the inferred rise time \citepalias{PaperI}. Thus, these four events are compelling candidates for possessing short-lived early flux excesses. If powered by ejecta--CSM collisions, they may represent less violent counterparts to 03fg-like events, featuring weaker early excesses and less significant ejecta deceleration, which preserves their typical light-curve stretch. Because the prominent early excesses in 03fg-like SNe likely originate from intense ejecta--CSM interactions following violent double WD mergers \citep{dimitriadisSN2021znyEarly2023,srivastavLuminousTypeIa2023}, we conclude that such extreme interactions remain rare among the normal SN\,Ia population.

SN\,2020adhi is particularly intriguing, displaying the longest adjusted rise time in our sample, which is comparable to those of many 03fg-like events. Among the identified long risers, it uniquely exhibits a near-linear rise ($\alpha_r\simeq1.0$). This combination is difficult to explain simply by superimposing an early flux excess onto an average SN\,Ia light curve \citepalias{PaperI}, and is accompanied by a high light-curve stretch ($x_1\simeq1.3$). Combined with the discrepancies observed against the \texttt{SALT2} model near peak, we propose that SN\,2020adhi is likely a misclassified peculiar SN\,Ia, potentially of the 03fg-like subclass. Although the absence of a high-quality spectrum near peak prevents definitive confirmation of its classification, it has a pre-peak spectrum\footnote{\url{https://ztfcosmo.in2p3.fr/target/ZTF20acyzsol}} that matches reasonably well with SN\,2006gz, a carbon-rich 03fg-like event \citep{hickenLuminousCarbonrichSupernova2007}.

Finally, we note that our search for early flux excesses is not meant to be exhaustive. The bias introduced by fitting an early flux excess with a power-law model scales with the amplitude \citepalias{PaperI}, and weak excesses can be hidden within the population-level scatter. For instance, \citet{deckersConstrainingTypeIa2022} conducted a systematic search for early flux excesses in the 2018 ZTF SN\,Ia sample and identified one candidate, SN\,2018cny (ZTF18aayjvve), that overlaps with our volume-complete sample \citep[see also][]{yeSearchingBumpsCosmological2024,rojas-bravoNotAllWho2026}. The flux excess in this event is less significant than those previously detected in the literature; indeed, it does not stand out in our analysis ($\alpha_r\simeq2.4$ and $t_\mathrm{rise} - t_{2.5\%, r}\simeq16.8$\,days). A calculation of the detection efficiency for such excesses would require a more detailed characterization of the population-level distribution of early-time parameters \citep[e.g.,][]{mageeDetectionEfficiencyType2022}.

Furthermore, the typical cadence and S/N of our sample make it challenging to constrain the prevalence of non-degenerate companions. Even from favorable viewing angles, a Roche-lobe-filling Solar-type or helium-star companion would produce only a relatively faint early excess \citep[$L_\mathrm{iso}\lesssim10^{41}\,\mathrm{erg\,s^{-1}}$, $\lesssim$1\% of the typical SN\,Ia bolometric peak luminosity;][]{kasenSEEINGCOLLISIONSUPERNOVA2009} that peaks predominantly in the ultraviolet (UV). Consequently, such an excess may not significantly alter the rise indices inferred from optical ZTF $g$- and $r$-band observations of an SN\,Ia at $z\simeq0.06$. Future high-cadence UV surveys will thus be essential for strictly constraining the presence of these companions.

\subsubsection{Shallow Risers: Extended Surface \Ni Distributions}\label{sec:discussion_anomalies_linear}

\begin{figure*}[ht!]
    \centering
    \includegraphics[width=\linewidth]{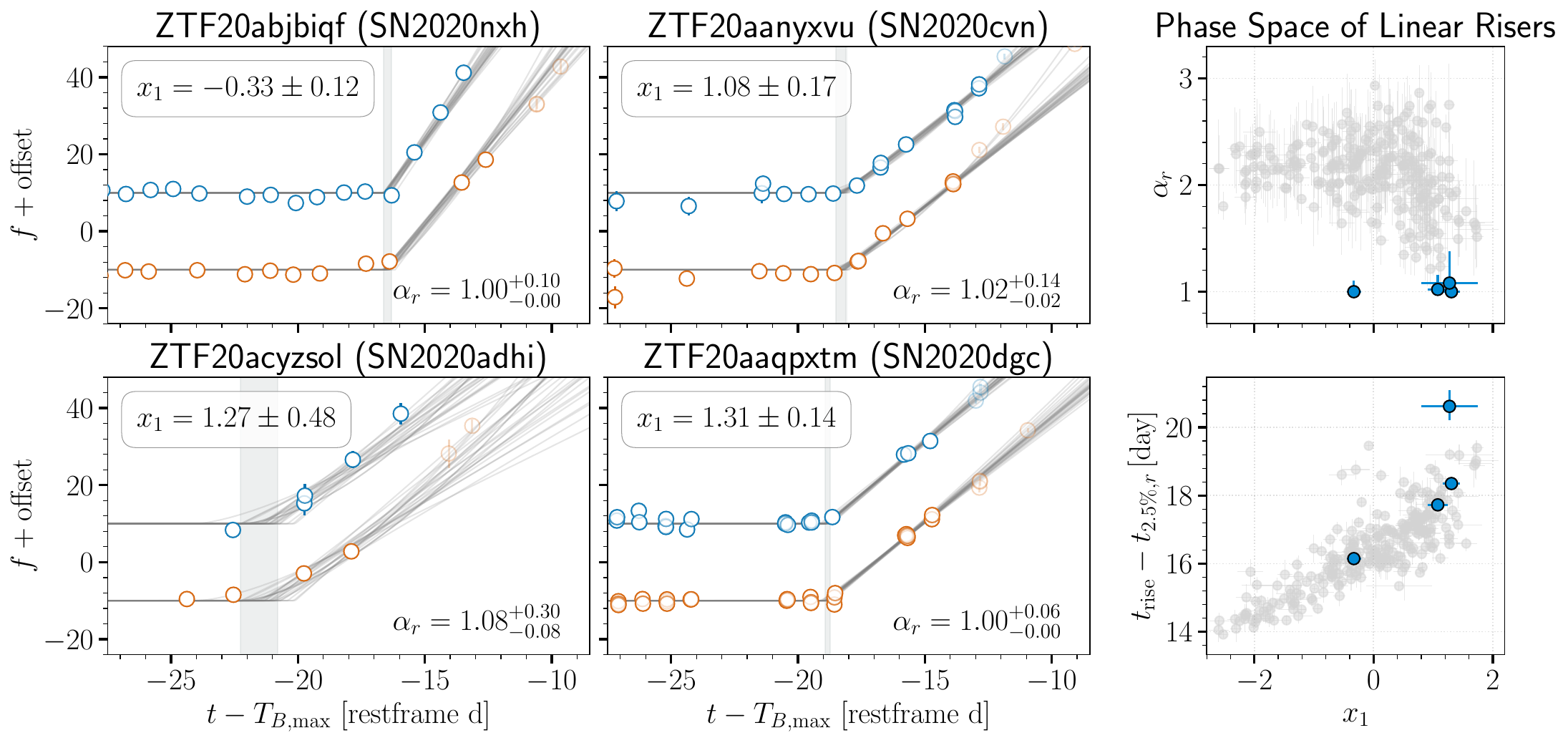}
    \caption{
        Light curve fitting results for the four SNe\,Ia consistent with a linear rise ($\alpha\simeq1$) in the volume-limited sample of ZTF DR2.  Plot formatting follows Figure~\ref{fig:ztf_anomalies_long}.
    }
    \label{fig:ztf_anomalies_linear}
\end{figure*}

Beyond 02cx-like events, several spectroscopically normal SNe\,Ia also display near-linear rises, such as SN\,2013dy \citep{zhengVERYYOUNGTYPE2013}, SN\,2014J \citep{zhengESTIMATINGFIRSTLIGHTTIME2014,goobarRISESN2014J2014}, and iPTF16abc \citep{millerEarlyObservationsType2018}. These objects are compelling candidates for harboring highly extended \Ni distributions within their outer ejecta.

Despite our strong prior forbidding $\alpha < 1$, we identify four SNe\,Ia with rise indices consistent with a linear evolution ($\alpha \simeq 1$; 16th percentile of the $\alpha_r$ posterior $<1.005$ and median $\alpha_r < 1.2$).
This behavior deviates from the main ZTF DR2 population. Figure~\ref{fig:ztf_anomalies_linear} highlights the early-time light curves of these events and their near-linear rises. SN\,2020adhi stands out once again (discussed in Section~\ref{sec:discussion_anomalies_long}). For the discussion below, we focus on the other three linear risers, which show typical rise times consistent with their respective $x_1$ values, and thus likely represent a distinct physical phenomenon.

Two of these SNe (SN\,2020dgc and SN\,2020cvn) are among the most intrinsically luminous and slowest-evolving events in our sample ($x_1 > 1$), whereas the other (SN\,2020nxh) exhibits a lower stretch ($x_1 \simeq -0.3$) and a proportionally shorter rise time. In addition to their anomalous rise indices, they exhibit persistently blue early colors ($\alpha_g - \alpha_r \simeq 0$). Spectroscopically, all three are consistent with overluminous 91T-like objects. Furthermore, with the exception of SN\,2020cvn, they display significant correlated residuals in their \texttt{SALT2} fits ($\texttt{fitprob} < 0.03$), indicating that the standard empirical model fails to fully capture their morphology even near maximum light. \citet{dr2_eex} also identified SN\,2020dgc (ZTF20aaqpxtm) as a candidate for hosting early flux excesses, though only when using the \texttt{SALT2} model as a baseline. Together, these discrepancies suggest that their unusual early-time behavior reflects a broader physical divergence from typical normal SNe\,Ia.

As shown in Figure~\ref{fig:ztf_anomalies_linear}, the linear risers with $x_1 \simeq 1$ are not strict outliers in the $\alpha$--$x_1$ plane, as there is no clear boundary separating these $\alpha \simeq 1$ events from the broader group. Although SN\,2020nxh initially appears to be an exception due to its low $x_1$, its spectroscopic similarity to 91T-like events and poor \texttt{SALT2} fit indicate that this measurement likely misrepresents its true light-curve shape. Indeed, the \texttt{SALT2} model significantly underestimates the broadness of its $g$-band emission.

We conclude that these 91T-like linear risers do not form an isolated sub-population. Instead, they occupy the extreme tail of early-time properties among the most luminous, slow-evolving normal SNe\,Ia. This corroborates spectroscopic evidence that 91T-like events represent a continuous extension of the normal population \citep[e.g.,][]{obrien1991TLikeTypeIa2024}. SN\,2020nxh is potentially a more exotic, low-stretch analog of the same underlying phenomenon.

The distinct dichotomy in the rise time and rise index between high-stretch normal/91T-like SNe and 03fg-like events indicates that early flux excesses do not share a universal physical origin. Although many high-stretch SNe\,Ia exhibit broadened early-time light curves and high peak luminosities, they are unlikely to be powered by the intense CSM interaction characteristic of 03fg-like events. This contrast points to distinct progenitor channels for the two populations.

Previous studies have noted an enhanced rate of early flux excesses among the high-stretch population \citep{jiangSurfaceRadioactivityInteractions2018,deckersConstrainingTypeIa2022,wuCommonOriginNormal2025,wuWFSTSupernovaeFirst2026,rojas-bravoNotAllWho2026}. 
Our analysis further suggests that among these high-stretch events, there is no strict dichotomy between those following a standard power-law rise and those exhibiting long-duration early flux excesses \citep[i.e., ``early-broad'' SNe\,Ia;][]{jiangSurfaceRadioactivityInteractions2018}. 
Instead, amplitudes of deviations from a standard power-law rise form a continuum. The prevalence of these deviations is independently supported by differences in the inferred $\alpha$ values across various truncation thresholds (Figure~\ref{fig:ztf_alpha_x1_corr}). This variation reflects a steepening of the rise over time, a feature most pronounced in the high-stretch population and likely driven by enhanced \Ni mixing in the outermost ejecta layers.

\subsection{Implications for Progenitor Demographics}

\begin{figure*}
    \centering
    \includegraphics[width=0.9\linewidth]{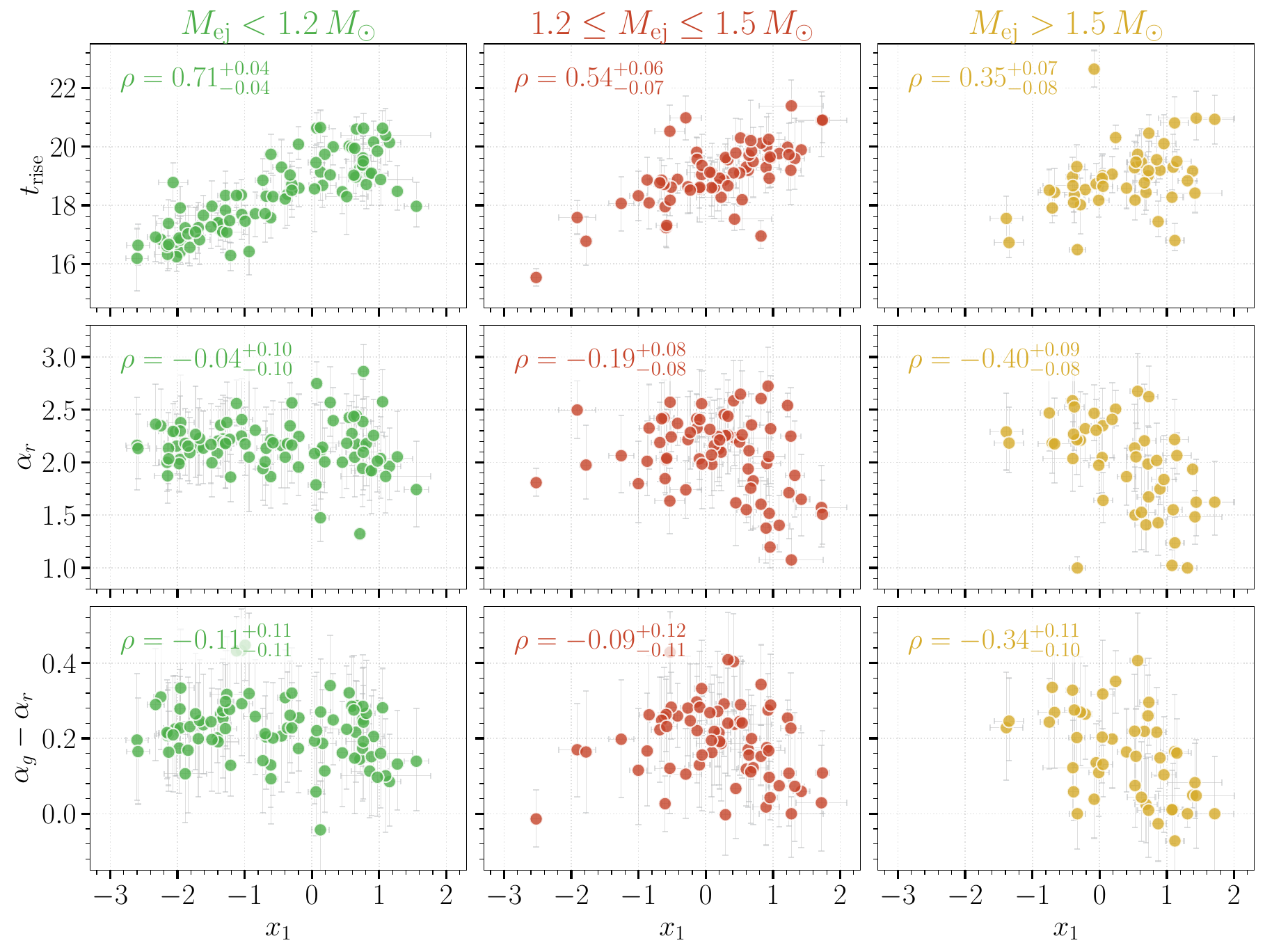}
    \caption{
        The early-time parameter ($t_\mathrm{rise}$, $\alpha_r$, $\alpha_g - \alpha_r$) correlations with \texttt{SALT2} $x_1$ are distinct for the sub-$\Mch$ (blue), near-$\Mch$ (green), and super-$\Mch$ (orange) populations defined in \citet{sarinLightcurveModelling22052026}. The posterior medians and 16--84th percentile ranges of the correlation coefficients $\rho$ are displayed for each relation.
    }
    \label{fig:ztf_t_rise_x1_corr_mej}
\end{figure*}

In Section~\ref{sec:corr_x1}, we demonstrate that the relationships between early-time parameters ($t_\mathrm{rise}$, $\alpha$, and $\alpha_g - \alpha_r$) and the full light-curve stretch $x_1$ are nonlinear, exhibiting a distinct break at some $x_\mathrm{brk}\gtrsim0$. 
Here, we explore the physical implications of these correlations and the apparent dichotomy between low- and high-stretch SNe\,Ia. Specifically, we investigate whether these observations point to distinct progenitor channels or can be explained by a continuous distribution of progenitor properties.

The light-curve stretch, broadly reflecting the diffusion timescale of the ejecta \citep{arnettTypeSupernovaeAnalytic1982}, is often interpreted as a proxy for the total ejecta mass. \citet{sarinLightcurveModelling22052026} recently performed systematic light-curve modeling for ZTF DR2 SNe\,Ia using the Arnett model, finding that the bimodal distribution of $x_1$ can be readily reproduced with a smooth, single-peaked distribution of ejecta masses spanning a wide range ($\sim$0.5--2\,$\Msun$). However, this continuous mass distribution does not necessarily preclude the existence of multiple distinct explosion mechanisms.

To understand how the inferred ejecta mass is associated with the early-time properties, we divide the ZTF DR2 sample into three groups, corresponding to the sub-$\Mch$ ($M_\mathrm{ej} < 1.2\,\Msun$), near-$\Mch$ ($1.2 \le M_\mathrm{ej} \le 1.5\,\Msun$), and super-$\Mch$ ($M_\mathrm{ej}>1.5\,\Msun$) populations defined in \citet{sarinLightcurveModelling22052026}. In Figure~\ref{fig:ztf_t_rise_x1_corr_mej}, we display the correlations between $x_1$ and the early-time parameters for the three populations.

The sub-$\Mch$ population features the tightest positive correlation with $t_\mathrm{rise}$ ($\rho\simeq0.7$), while no significant correlations are found with either $\alpha_r$ or $\alpha_g - \alpha_r$. Conversely, the super-$\Mch$ population displays mild yet significant anticorrelations between $x_1$ and $\alpha_r$ ($\rho\simeq-0.4$), as well as between $x_1$ and $\alpha_g - \alpha_r$ ($\rho\simeq-0.3$). These anticorrelations, detected at the 3--4$\sigma$ significance level, drive the trends observed in the high-stretch population. Additionally, the correlation between $x_1$ and $t_\mathrm{rise}$ in the super-$\Mch$ group is considerably weaker than in the sub-$\Mch$ population with more prominent scatter. Finally, the near-$\Mch$ population exhibits intermediate behavior, showing a moderate correlation between $x_1$ and $t_\mathrm{rise}$ but lacking significant correlations with $\alpha_r$ or $\alpha_g - \alpha_r$.

Focusing exclusively on the high-stretch population, we observe distinct early-time properties among the three groups. Despite containing comparable numbers of SNe with $0.5 < x_1 < 1.5$ (23, 25, and 23 for the sub-$\Mch$, near-$\Mch$, and super-$\Mch$ populations, respectively), the median $\alpha_r$ of SNe within this interval decreases systematically with increasing inferred $M_\mathrm{ej}$, dropping from $2.18\pm0.11$ (sub-$\Mch$) to $1.93\pm0.14$ (near-$\Mch$) and $1.79\pm0.11$ (super-$\Mch$). Similarly, the median $\alpha_g - \alpha_r$ values decline progressively from $0.20\pm0.04$ to $0.16\pm0.04$ and $0.10\pm0.04$. This trend is evident in Figure~\ref{fig:ztf_t_rise_x1_corr_mej}, which reveals clusters of $x_1\simeq1$ events with $\alpha_r\simeq1.5$ and $\alpha_g - \alpha_r\simeq 0$ in the near- and super-$\Mch$ groups, but absent in the sub-$\Mch$ group. Although the median $t_\mathrm{rise}$ remains consistent across all three, these variations in rise indices and early color evolution reinforce the conclusion drawn in Section~\ref{sec:corr_x1_t_rise} that the stretch alone cannot fully capture early-time diversity. Instead, the inferred ejecta mass likely encodes more fundamental physical details regarding the outer ejecta layers.

Nevertheless, we caution that the $M_\mathrm{ej}$ estimates from \citet{sarinLightcurveModelling22052026} are subject to systematic uncertainties, as their modeling assumes spherical symmetry and a well-mixed \Ni distribution. Asymmetric explosions can naturally introduce $\sim$0.1--0.2\,$\Msun$ scatter in the $M_\mathrm{ej}$ estimation due to viewing-angle effects in observables \citep{kasenTimedependentMonteCarlo2006,townsleyDoubleDetonationsThin2019,shenMultidimensionalRadiativeTransfer2021}. More fundamentally, fitting a confined \Ni distribution within the Arnett framework (which assumes instantaneous mixing) inflates the diffusion timescale, thereby systematically overestimating $M_\mathrm{ej}$ \citep{khatamiPhysicsLuminousTransient2019}. 
Moreover, if, as suggested by our findings, the degree of \Ni mixing varies across the population, with the high-stretch population generally featuring more extended \Ni distributions, this upward bias will vary on a case-by-case basis. This further complicates the interpretation of the inferred $M_\mathrm{ej}$ values. 
Consequently, although the inferred super-$\Mch$ population exhibits fundamental differences in the light curves compared to the rest of the sample, their progenitors are not necessarily super-$\Mch$ in reality. The relative ordering of the $M_\mathrm{ej}$ values among the three groups should, however, be generally robust.

Self-consistent modeling of early-time multi-band light curves requires realistic treatments of both the \Ni distribution and the ejecta density profile. A smooth variation in \Ni-mixing degrees should only produce a continuous trend between early emission and stretch; the observed break therefore suggests a more complex physical reality. This could indicate distinct explosion mechanisms, or alternatively, if a single mechanism dominates, an abrupt structural change in the ejecta at a critical progenitor condition or prominent ejecta asymmetries producing complicated viewing-angle effects.

To evaluate these possibilities, we examine the predicted correlations from two detailed theoretical frameworks: (i) near-$\Mch$ explosions spanning a grid of density profiles and degrees of \Ni mixing \citep{mageeDetermining56Ni2020}, and (ii) sub-$\Mch$ double-detonation models featuring varying WD masses and viewing angles \citep{shenMultidimensionalRadiativeTransfer2021}. By leveraging the flexibility of these models to produce a diverse range of light-curve shapes and early color evolution trends, we assess whether a single underlying mechanism could reproduce the observed parameter space.

The relations between early-time parameters and full light-curve shapes here serve as diagnostics for theoretical models. However, the \texttt{SALT2} stretch $x_1$ lacks a physical counterpart in synthetic light curves generated by radiative transfer simulations. To bridge this gap, we propose the adjusted rise time $t_\mathrm{rise} - t_{\mathrm{2.5}\%, r}$\footnote{As shown in Figure~\ref{fig:ztf_t_rise_x1_corr_adjusted}, this diagnostic is relatively insensitive to the choice of X\%.} as a physically motivated, easily measurable proxy for $x_1$ (see Section~\ref{sec:corr_x1_t_rise}) in both real and synthetic datasets. Its relations with $\alpha_r$ and $\alpha_g - \alpha_r$ exhibit a distinct break at $\sim$16--17\,days (corresponding to $x_1\simeq0$--0.5), above which anticorrelations with both parameters emerge.

\subsubsection{Near-$\Mch$ Progenitors: Fluctuations in Density Profiles and \Ni Distributions}\label{sec:discussion_models_turtls}

\begin{figure*}
    \centering
    \includegraphics[width=0.9\linewidth]{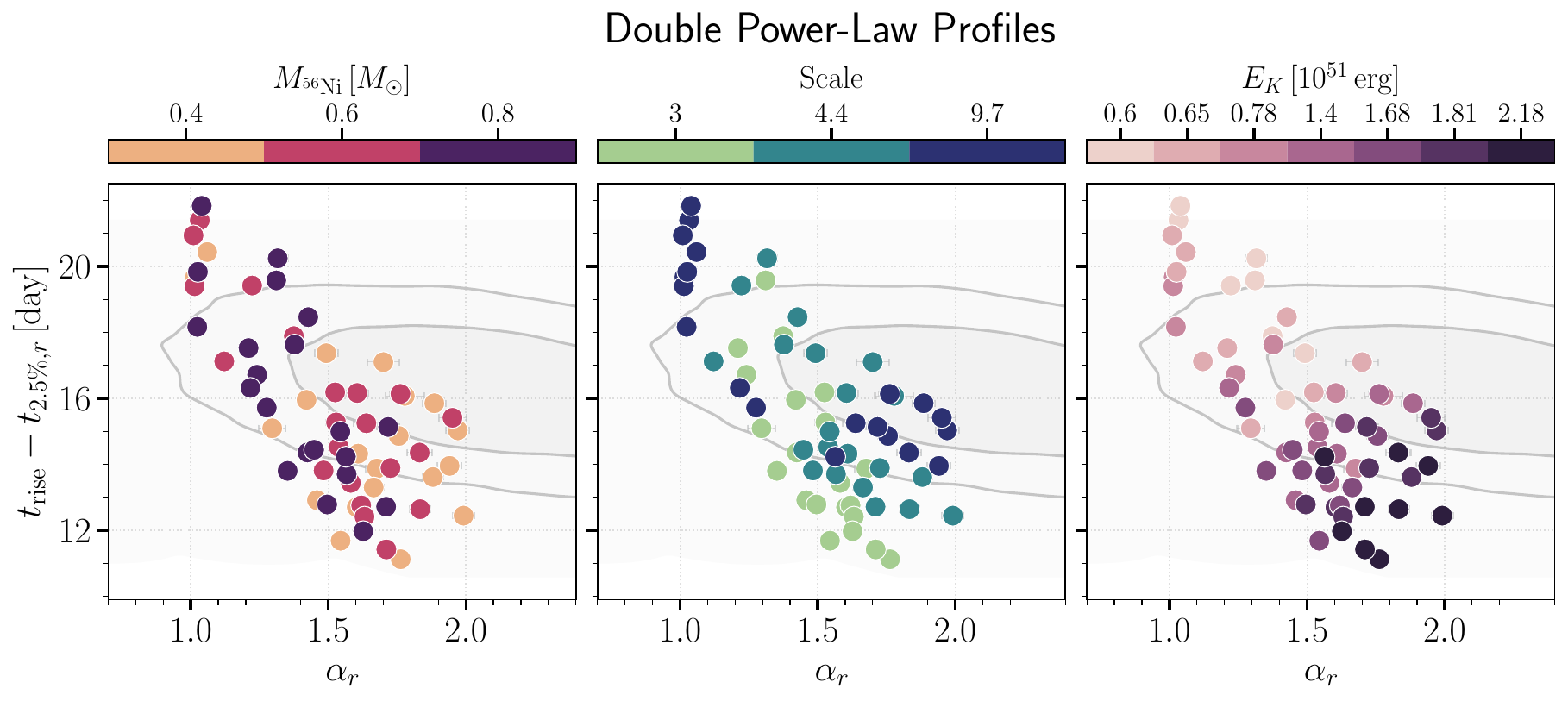}
    \includegraphics[width=0.9\linewidth]{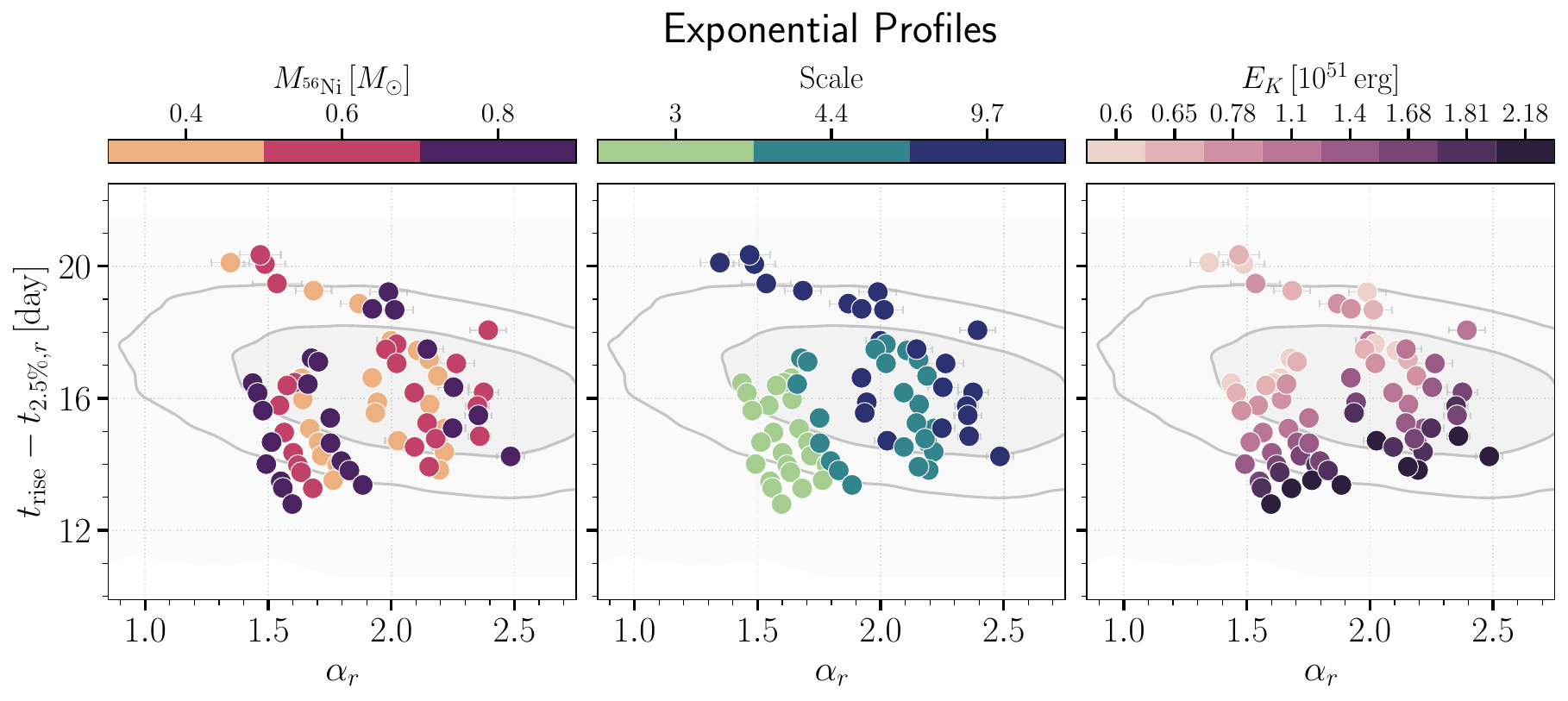}
    \caption{
        The relation between the rise index $\alpha_r$ and the adjusted rise time $t_\mathrm{rise} - t_{2.5\%, r}$ for the TURTLS models with double power-law (top) and exponential (bottom) density profiles. Each point represents a single synthetic light curve, color-coded by the progenitor parameters: total \Ni mass ($M_\mathrm{^{56}Ni}$), mixing scale factor, and kinetic energy ($E_K$). 
        Overplotted are the 68\% and 95\% contours of the predictive distribution from the population prior applied in the modeling.
        The models predict a wide diversity of early-time behaviors, but generally fail to reproduce the observed correlations with the adjusted rise time.
    }
    \label{fig:turtls_alpha}
\end{figure*}

\begin{figure*}
    \centering
    \includegraphics[width=0.9\linewidth]{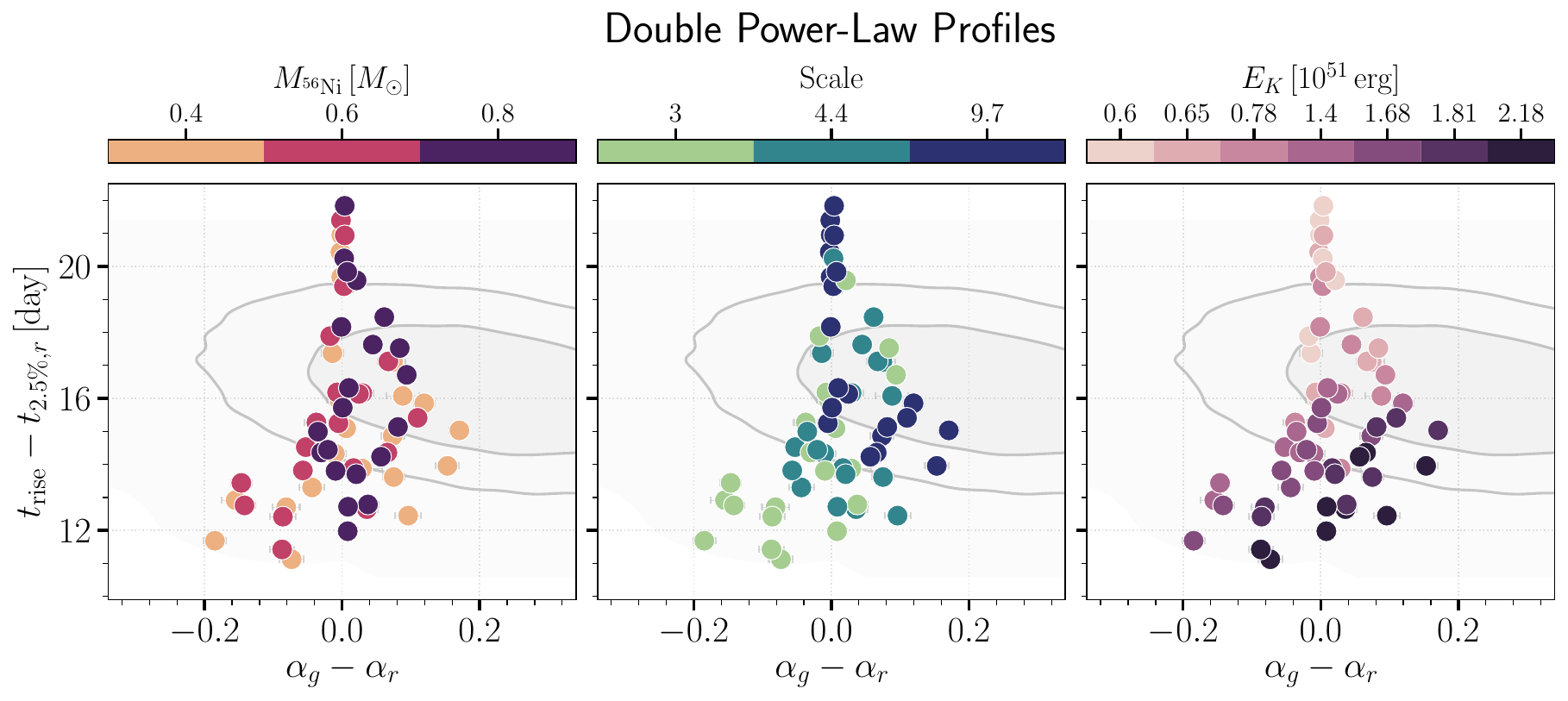}
    \includegraphics[width=0.9\linewidth]{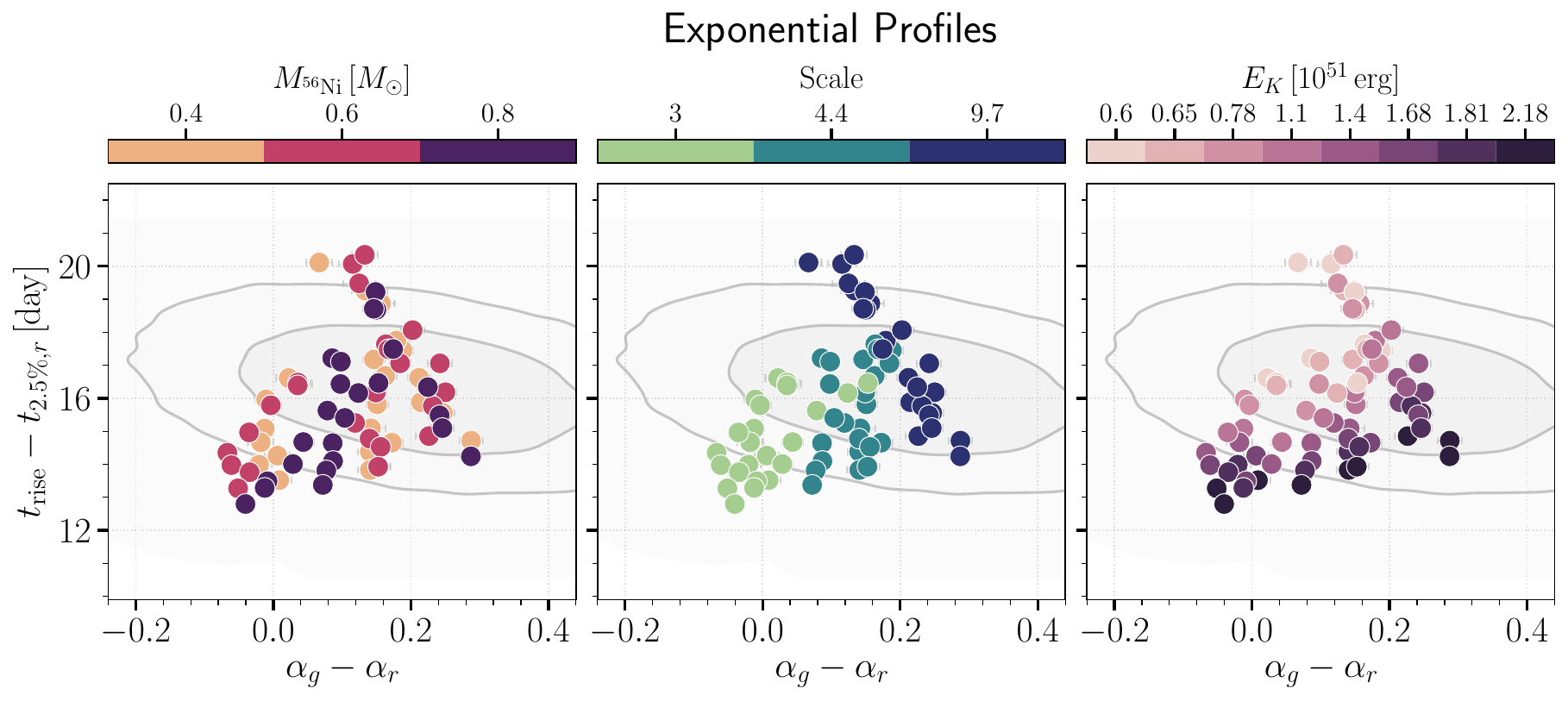}
    \caption{
        The relation between the early-time color evolution index $\alpha_g - \alpha_r$ and the adjusted rise time $t_\mathrm{rise} - t_{2.5\%, r}$ for the same TURTLS models as in Figure~\ref{fig:turtls_alpha}. No significant correlations are found between $\alpha_g - \alpha_r$ and the adjusted rise time.
    }
    \label{fig:turtls_alpha_g_r}
\end{figure*}

\citet{mageeDetermining56Ni2020} conducted a grid search for near-$\Mch$ progenitors, varying the ejecta density profiles (functional forms and kinetic energy) and \Ni distributions (total \Ni mass and a scale factor governing outward mixing). They generated multi-band synthetic light curves using the 1D radiative transfer code TURTLS \citep{mageeModellingEarlyTime2018}. Although the parameter combinations in this extensive grid are not strictly tied to specific, self-consistent explosion models, they provide a valuable baseline for exploring the wide array of light-curve morphologies and early-time behaviors that can arise from near-$\Mch$ explosions.

For a statistically robust comparison between the models and observations, we generate synthetic light curves for an extensive grid of models that broadly represent the diversity of density profiles and \Ni distributions. Specifically, we consider two sets of density profiles (double power-law and exponential), and restrict the analysis to models with a mixing scale factor of $<$10 \citep[representing a reasonable degree of mixing;][]{mageeDetermining56Ni2020} and a kinetic energy of $E_K > 0.5\times10^{51}$\,erg (yielding realistic rise times of $\lesssim$20\,days). 
For each model, we generate 50 synthetic light curves, drawn from a uniform volumetric distribution up to $z=0.06$. 
The TURTLS models provide synthetic light curves with a 1-day cadence in absolute magnitudes, which we convert to fluxes (without $K$-corrections). To assign realistic flux uncertainties, we randomly draw sky noise from the ZTF pointings catalog adopted in \texttt{REDBACK} \citep{sarinREDBACKBayesianInference2024} and add a flat 2\% of the model source flux in quadrature to account for calibration uncertainties. We also pad the pre-explosion light curves with zero-flux data points, sampled at a 2-day cadence with randomly drawn sky noise out to 50\,days before $B$-band maximum. Finally, the fluxes are normalized to the peak flux in their respective bands (Section~\ref{sec:ztf}) and fit to a power-law rise using the population priors inferred from the ZTF DR2 sample to ensure a direct comparison with the observed correlations.

Even for the same underlying model, different noise realizations yield a wide range of inferred parameters, particularly for the rise index $\alpha$. To robustly estimate the mean of each parameter across the 50 synthetic light curves, we adopt the weighted bootstrap method from \citetalias{PaperI}. This approach reweights the parameters from each light curve by the inverse variance of their posteriors, downweighting poorly constrained noise realizations to produce reliable posterior draws for the mean.

The resulting distributions are shown in Figures~\ref{fig:turtls_alpha} and \ref{fig:turtls_alpha_g_r}, with the median and 16th--84th percentile ranges of $\alpha_r$ and $\alpha_g - \alpha_r$ plotted against the adjusted rise time for each model.
TURTLS models predict a wide diversity of early-time behaviors, despite the regularization imposed by the population priors (the prior predictives are overplotted as contours). The choice of density profile significantly influences the early-time behavior, with double power-law profiles generally yielding shallower rises compared to exponential profiles.
For both density profiles, models with a higher degree of \Ni mixing (lower scale factors) exhibit shallower rises (lower $\alpha$) and bluer early colors (low $\alpha_g - \alpha_r$), while those with lower kinetic energy systematically produce longer rise times. No significant correlations are found between the early-time parameters and the \Ni mass. 

A notable exception occurs in models combining a relatively confined \Ni distribution (scale factor $= 9.7$) with low kinetic energy ($E_K\lesssim 0.8\times10^{51}$\,erg), which also yield low $\alpha$ values. This combination creates a prolonged dark phase where early diffusing photons remain too faint to detect: the typical delay of the inferred first light time $t_\mathrm{fl}$ relative to the explosion is $\lesssim$1\,day, but can last up to $\gtrsim$3\,days in these specific models. Consequently, the inferred $\alpha$ is driven by later epochs after the photosphere recedes into the \Ni-rich layers. 
Notably, the long-dark-phase models with double power-law density profiles could produce $\alpha\simeq1$, offering an alternative explanation for the linear risers observed in our sample (Section~\ref{sec:discussion_anomalies_linear}) without requiring highly extended \Ni distributions. However, these models also predict extraordinarily long rises ($t_\mathrm{rise} - t_{2.5\%, r} \gtrsim 20$\,days) that are not observed in our normal SN\,Ia sample.

Despite this broad diversity, the models fail to reproduce the observed correlations between early-time parameters and our light-curve stretch proxy. Double power-law models predict an anticorrelation between the adjusted rise time and $\alpha_r$ across the entire population, which, however, lacks a structural break. Furthermore, these models show a positive correlation between $\alpha_g - \alpha_r$ and the adjusted rise time for $t_\mathrm{rise} - t_{2.5\%, r}\lesssim16$\,days, which contradicts the lack of correlation observed in the low-stretch population.\footnote{The apparent break in the $\alpha_g - \alpha_r$ relation for double power-law models is an artifact due to the presence of linear risers ($\alpha\simeq1$); because our prior forbids $\alpha<1$, the $\alpha_g - \alpha_r$ values for these models are artificially driven to zero.} Meanwhile, models with exponential density profiles yield no significant correlations between any early-time parameters and the adjusted rise time.

Nevertheless, the discrepancies between the TURTLS models and the observed correlations should be interpreted with caution. The TURTLS models are not derived from self-consistent explosion simulations, and their parameter space may not fully capture the range of physical conditions present in real SNe\,Ia, particularly the density profiles, which we have shown to significantly influence the early-time behavior. Moreover, the 1D nature of these models inherently limits their ability to reproduce the complex multi-dimensional effects arising from asymmetries in the ejecta structure, which may play a crucial role in shaping the early emission.

\subsubsection{Sub-$\Mch$ Double-detonations: WD Masses and Viewing Angle Effects}\label{sec:discussion_models_shen}

\begin{figure*}
    \centering
    \includegraphics[width=0.67\linewidth]{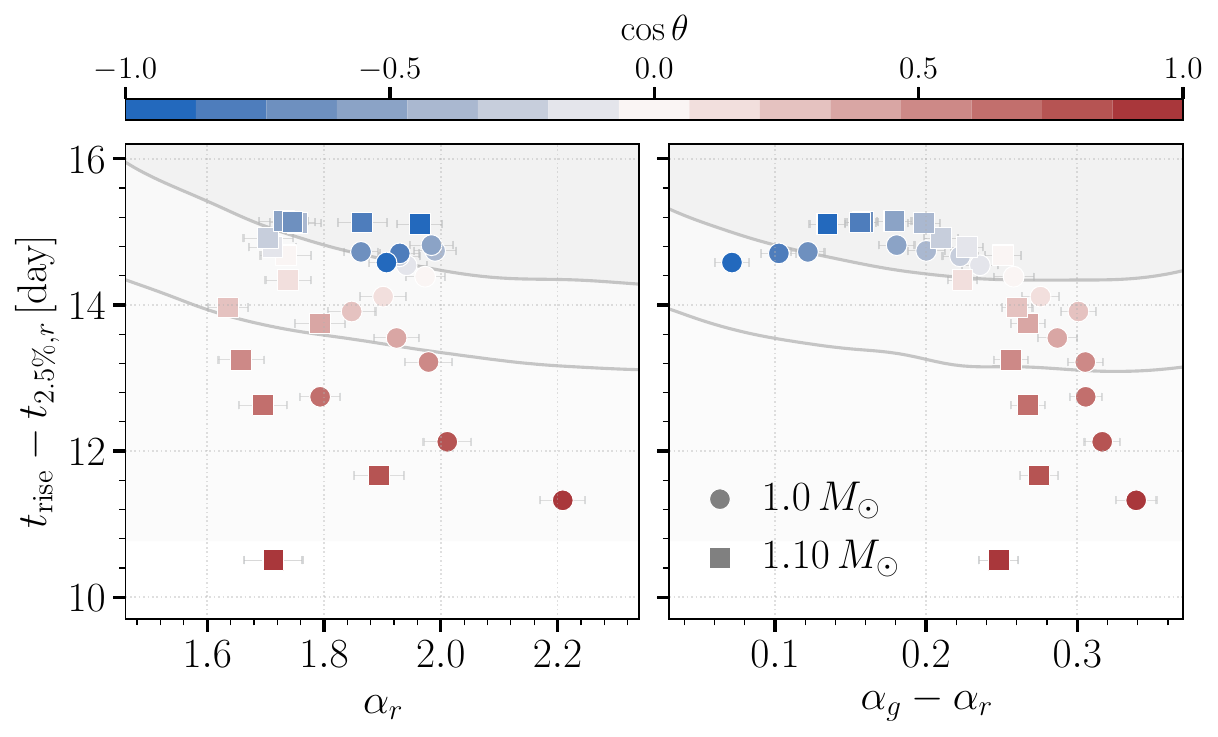}
    \caption{
        The relationship between the adjusted rise time $t_\mathrm{rise} - t_{2.5\%, r}$ with the rise index $\alpha_r$ (\textit{left}) and color evolution index $\alpha_g - \alpha_r$ (\textit{right}) for the sub-$\Mch$ double-detonation models from \citet{shenMultidimensionalRadiativeTransfer2021}. Each point represents the median value and 16th--84th percentile range of the inferred parameter mean for 100 synthetic light curves at a specific viewing angle, which is color-coded. Overplotted are the 68\% and 95\% contours of the population prior predictives. These models predict a break in the correlation between the adjusted rise time and $\alpha_g - \alpha_r$, driven by the inherent asymmetric \Ni distribution and density structure characteristic of the double-detonation mechanism.
    }
    \label{fig:shen}
\end{figure*}

Double-detonation explosions of sub-$\Mch$ WDs have long been proposed as a promising progenitor channel for normal SNe\,Ia \citep{shenUNSTABLEHELIUMSHELL2009,finkDoubledetonationSubChandrasekharSupernovae2010,kromerDOUBLEDETONATIONSUBCHANDRASEKHARSUPERNOVAE2010,gronowSNeIaDouble2020,gronowDoubleDetonationsSubMCh2021}. In this scenario, a detonation in a thin helium shell on the WD surface triggers a secondary detonation in the underlying carbon-oxygen core, producing an SN\,Ia-like explosion. The resulting ejecta structure is inherently asymmetric due to the off-center ignition of the helium shell, leading to strong viewing-angle effects on the light curves and spectra \citep[as found in SN\,2022joj;][]{liuSN2022jojPeculiar2023}.

\citet{shenMultidimensionalRadiativeTransfer2021} performed 2D radiative transfer calculations for a grid of sub-$\Mch$ double-detonation models, utilizing varying WD and helium shell masses from \citet{boosMultidimensionalParameterStudy2021}. These simulations predict a rich diversity of early-time light-curve shapes and color evolution trends driven by the interplay between WD mass, helium shell mass, and viewing angle. Varying these progenitor parameters reproduces a wide range of peak luminosities and light-curve morphologies that are broadly consistent with the observed luminosity--stretch relation of the overall SN\,Ia population.

To test whether these sub-$\Mch$ models can reproduce the observed empirical correlations between early-time parameters and stretch, we consider two representative thin-shell models. Both models feature a low base helium shell density of $\rho=2\times10^5\,\mathrm{g\,cm^{-3}}$, ensuring that the iron-group elements synthesized during the shell detonation have minimal impact on the observables. We select progenitor masses of $1.0\,\Msun$ and $1.1\,\Msun$, yielding total \Ni masses of $0.495\,\Msun$ and $0.747\,\Msun$, respectively, which produce peak luminosities comparable to those of normal SNe\,Ia. For both models, we generate 100 synthetic light curves for each of the 15 viewing angles ($\theta$, measured from the helium-shell detonation location), following the same procedure as described for the TURTLS models in Section~\ref{sec:discussion_models_turtls}.
Finally, we fit these synthetic light curves using the same population priors extracted by modeling the ZTF DR2 sample.

Figure~\ref{fig:shen} shows the inferred $\alpha_r$ and $\alpha_g - \alpha_r$ as a function of the adjusted rise time. Again, we display the median and 16th--84th percentile range of the parameter means across 100 realizations of each model and viewing angle.
The viewing angle effects drive a wide diversity in early-time behavior, whereas the progenitor mass only has a mild influence on the rise index and color evolution: the more massive model generally produces slightly shallower rises and weaker color evolution, particularly for viewing angles $\theta<\pi/2$ (on the hemisphere where the helium-shell detonation occurs). Both WD masses yield similar adjusted rise times ($\sim$10--15\,days), which are systematically shorter than those observed in the ZTF DR2 sample,\footnote{Even with a full non-LTE treatment, the epoch of $B$-band maximum light for these models is delayed by no more than $\sim$1 day relative to the LTE calculation \citep{boosNonLTESyntheticObservables2025}.} aligning only with the low-stretch end of the observed population.

Despite discrepancies in overall timescales, both models exhibit a strong break in the correlation between the adjusted rise time and $\alpha_g - \alpha_r$. For viewing angles $\theta > \pi/2$, the adjusted rise time plateaus near $\sim$15\,days, while $\alpha_g - \alpha_r$ decreases dramatically as the sightline approaches the antipode of the detonation, corresponding to bluer early colors. Conversely, when $\theta < \pi/2$, $\alpha_g - \alpha_r$ remains consistently high, but the adjusted rise time decreases as the sightline aligns more directly with the detonation.
This trend is driven by the asymmetric \Ni distribution inherent to the double-detonation mechanism. Because the \Ni-rich core is offset toward the hemisphere opposite the initial ignition, sightlines closer to the detonation exhibit not only lower peak luminosity \citep{shenMultidimensionalRadiativeTransfer2021}, but also shorter rise times and a pronounced red$\rightarrow$blue evolution. In contrast, sightlines closer to the \Ni-rich core yield longer rise times and persistently blue early colors. 

The correlation between the adjusted rise time and $\alpha_r$ is much weaker, with a mild anticorrelation for $\theta < \pi/2$. The $\alpha_r$ values are generally low ($\lesssim$2) across all viewing angles, despite the population priors pulling toward a higher value. The typical dark phase duration for these models is relatively short ($\lesssim$1\,day) across all viewing angles, and thus does not significantly influence the inferred $\alpha_r$ values.

In summary, the sub-$\Mch$ double-detonation models demonstrate that a single explosion mechanism characterized by significant ejecta asymmetry provides one possible explanation for the broken correlations between the adjusted rise time and early color evolution, although the break is absent in the relation between the adjusted rise time and the rise index. Furthermore, these models underpredict the rise times, a persistent challenge when attempting to reconcile double-detonation scenarios with the macroscopic properties of normal SNe\,Ia.

\section{Conclusions}\label{sec:conclusions}

In this second paper of our series on the early-time light curves of SNe\,Ia, we performed statistical inference on a volume-complete sample of 972 normal SNe\,Ia from the ZTF DR2, an order of magnitude larger than any previously used in similar studies. 
We leveraged the hierarchical Bayesian framework developed in \citetalias{PaperI} to incorporate sparsely sampled light curves, which avoids the selection bias induced by applying strict quality cuts. This approach yielded the first comprehensive characterization of the early-time parameter space for a large, homogeneous sample of SNe\,Ia, establishing robust population-level constraints on the key observables: the rise time ($t_\mathrm{rise}$; $\mu=18.55\pm0.08$\,days, $\sigma=1.42\pm0.07$\,days), rise index ($\alpha$; $\mu=2.10\pm0.04$, $\sigma=0.48\pm0.03$ in ZTF $r$), and $g-r$ color evolution ($\alpha_g - \alpha_r$; $\mu=0.20\pm0.02$, $\sigma=0.17\pm0.02$). There is a weak positive correlation between $t_\mathrm{rise}$ and $\alpha$, whereas $\alpha_g$ and $\alpha_r$ are almost perfectly correlated.

Focusing on 272 SNe\,Ia with high-quality early-time coverage, we identified a dichotomy in how early-time parameters correlate with the light-curve stretch (\texttt{SALT2} $x_1$). A break occurs near $x_\mathrm{brk}\simeq 0.2$--0.5: for high-stretch events, both $\alpha$ and $\alpha_g - \alpha_r$ display significant anticorrelations with $x_1$, whereas the low-stretch population lacks such correlations. Additionally, $t_\mathrm{rise}$ increases more slowly with $x_1$ for high-stretch events. These relations are subject to significant intrinsic scatter, and should be interpreted as broad trends rather than deterministic relations. Incorporating $x_1$ directly into the hierarchical framework in future work will place tighter constraints on this structural break.

Beyond the normal population, peculiar subgroups occupy distinct regions in the parameter space, implying diverse outer ejecta properties. Overluminous 03fg-like events display significantly longer $t_\mathrm{rise}$ relative to normal SNe\,Ia, whereas subluminous 02cx-like and 91bg-like events exhibit notably shorter $t_\mathrm{rise}$. Furthermore, the distinct rise profiles of these peculiar subclasses, steeper for 03fg-like and shallower for 02cx-like events, independently support violent mergers and partial deflagration as their respective physical origins.

Within the normal sample, we identified multiple objects with anomalous early-time properties, including SNe\,Ia with exceptionally long rise times relative to their stretch, and others with near-linear rises ($\alpha \simeq 1$). The extended rise times for the outliers with a normal stretch ($x_1\simeq0$) likely arise from short-lived early flux excesses that bias the inferred time of first light, potentially driven by CSM interaction.
The linear risers, predominantly high-stretch 91T-like SNe, do not form an isolated subclass. Instead, they likely represent the tail of a continuous distribution of early-time properties among the most luminous, slow-evolving normal SNe\,Ia, suggesting a diverse degree of enhanced \Ni mixing appears prevalent among the high-stretch population. Whether this enhanced mixing drives the observed anticorrelation between $\alpha$ and $x_1$, or instead only contributes to the overall scatter, merits further investigation.

These empirical findings offer new constraints on progenitor scenarios. We compared our results to detailed theoretical models, including 1D near-$\Mch$ models with varying density and \Ni profiles \citep{mageeDetermining56Ni2020}, and 2D sub-$\Mch$ double-detonations with varying progenitor masses and viewing angles \citep{shenMultidimensionalRadiativeTransfer2021}. While these models predict a wide diversity of light-curve shapes, they generally fail to reproduce the specific correlations with stretch. 
This discrepancy is likely due to a combination of two major factors: (i) unrealistic \Ni and ejecta density profiles arising from simplified parameterizations or limitations in hydrodynamic simulations, and (ii) radiative transfer calculations that struggle to capture the multi-dimensional effects in the extremely optically thick SN ejecta shortly after explosion. 
Multi-dimensional explosion models with more realistic progenitor setups are necessary to fully explain the early-time diversity of SNe\,Ia. Furthermore, utilizing a large sample of bolometric light curves constructed with multi-band, high-cadence photometry can significantly reduce the systematics introduced by radiative transfer modeling, providing more direct constraints on the underlying physical parameters.

Spectroscopic follow-up across extended phases provides crucial diagnostics for understanding the physical origins of the observed correlations and anomalies. For instance, pre-peak spectra reveal the ejecta velocity structure and chemical composition of the outer layers, while nebular-phase spectra probe the innermost ejecta structure and nucleosynthetic yields. Connecting early-time photometric properties to these spectroscopic diagnostics yields deeper insights into the explosion mechanisms shaping the observed diversity. Additionally, host-galaxy properties uniquely constrain progenitor demographics and delay-time distributions, independently informing the progenitor channels. In future work of this series, we will systematically explore these connections across the ZTF SN\,Ia sample.

The observed nonlinear correlations expose a multimodal parameter space that our current single multivariate Gaussian population prior cannot naturally capture. Future hierarchical models should incorporate more flexible population priors, such as Gaussian mixture models or non-parametric frameworks. Although these approaches require larger sample sizes, the rapidly expanding ZTF dataset ($\sim$$10^4$ SNe\,Ia in the upcoming DR3) and next-generation surveys---including Rubin Observatory's Legacy Survey of Space and Time \citep{ivezicLSSTScienceDrivers2019}, the Roman Telescope's High-Latitude Time-Domain Survey, and high-cadence ground-based programs like the La Silla Schmidt Southern Survey \citep{millerSillaSchmidtSouthern2025}---will readily supply the data necessary to drive these advanced analyses.

\software{\texttt{ArviZ} \citep{martinArviZModularFlexible2026}, \texttt{Astropy} \citep{astropycollaborationAstropyCommunityPython2013,astropycollaborationAstropyProjectBuilding2018,astropycollaborationAstropyProjectSustaining2022}, \texttt{JAX} \citep{jax2018github}, \texttt{NumPyro} \citep{binghamPyro2019, phanComposableEffectsFlexible2019}, \texttt{REDBACK} \citep{sarinREDBACKBayesianInference2024}}

\begin{acknowledgments}
We thank Mark Magee and Ken Shen for providing the synthetic light curves from their models. We thank Vicky Kalogera and Wen-fai Fong for the fruitful discussions.
C.L. and~A.A.M.~are supported by DoE award \#\,DE-SC0025599, while A.A.M.~is also supported by Cottrell Scholar Award \#\,CS-CSA-2025-059 from Research Corporation for Science Advancement.
NS acknowledges support from the Kavli Foundation. 
K.M. and~T.E.M.B.~acknowledge funding from Horizon Europe ERC grant no. 101125877.
L.G. acknowledges financial support from CSIC, MCIN and AEI 10.13039/501100011033 under projects PID2023-151307NB-I00, PIE 20215AT016, CEX2020-001058-M, and by the MaX-CSIC Excellence Award MaX4-SOMMA-ICE.
Y.-L.K. was supported by the Lee Wonchul Fellowship, funded through the BK21 Fostering Outstanding Universities for Research (FOUR) Program (grant No. 4120200513819) and the National Research Foundation of Korea(NRF) grant funded by the Korea government(MSIT) (RS-2022-NR070872; RS-2022-NR070525; RS-2026-25473561).

This work is based on observations obtained with the Samuel Oschin Telescope 48-inch and the 60-inch Telescope at the Palomar Observatory as part of the Zwicky Transient Facility project. ZTF is supported by the National Science Foundation under Grant No. AST-1440341 and a collaboration including Caltech, IPAC, the Weizmann Institute of Science, the Oskar Klein Center at Stockholm University, the University of Maryland, the University of Washington, Deutsches Elektronen-Synchrotron and Humboldt University, Los Alamos National Laboratories, the TANGO Consortium of Taiwan, the University of Wisconsin at Milwaukee, and Lawrence Berkeley National Laboratories. Operations are conducted by COO, IPAC, and UW.

This work used computing resources provided by Northwestern University and the Center for Interdisciplinary Exploration and Research in Astrophysics (CIERA). This research was supported in part through the computational resources and staff contributions provided for the Quest high performance computing facility at Northwestern University which is jointly supported by the Office of the Provost, the Office for Research, and Northwestern University Information Technology.
\end{acknowledgments}

\appendix
\section{Impact of Ignoring $K$-corrections in Light-curve Modeling} 
\label{sec:mock_kcorr}
$K$-corrections are usually ignored in early-time SN\,Ia analysis due to the lack of a robust early-time spectrophotometric sample, which can introduce a systematic uncertainty in the rise index $\alpha$ of $\sim$0.1 \citep{firthRisingLightCurves2015}. Consequently, modeling an SN\,Ia sample with a wide redshift distribution without accounting for $K$-corrections could induce spurious scatter and correlations in the inferred light-curve parameters.

To quantify this effect, we design simulations based on the spectrophotometric time series of SN\,2011fe \citep{pereiraSpectrophotometricTimeSeries2013}, the gold-standard normal SN\,Ia with the most complete early-time spectral coverage with accurate flux calibration. We generate synthetic light curves of SN\,2011fe as observed at different redshifts in the ZTF $g$ and $r$ bands, using the time-dependent spectral energy distribution (SED) from \citet{pereiraSpectrophotometricTimeSeries2013}, and then fit these synthetic light curves using the same hierarchical Bayesian framework. 

We first model synthetic light curves of SN\,2011fe at two fixed redshifts: $z=0.01$ (where $K$-corrections are minimal) and $z=0.06$ (the bound of the ZTF volume-complete sample), fitting them up to 40\% of peak flux with the simple power-law model. For both sets of 300 synthetic SNe, the inferred scatter in $\alpha$ and $t_\mathrm{rise}$ is consistent with zero, reflecting the single underlying SED. However, the inferred $\alpha_g$ at $z=0.06$ is systematically larger than at $z=0.01$ by $0.10\pm0.03$, while differences in $\alpha_r$ ($0.01\pm0.03$) and $t_\mathrm{rise}$ ($0.06\pm0.07$\,days) are not statistically significant. This bias in the $g$ band is driven by broad, rapidly evolving absorption features in the rest-frame blue optical spectrum of SN\,2011fe. Using a truncation threshold of 30\%, where the early SED is more strongly shaped by these features, the effect becomes more pronounced: the difference in $\alpha_g$ increases to $0.30\pm0.05$, and discrepancies in $\alpha_r$ ($0.18\pm0.15$) and $t_\mathrm{rise}$ ($0.46\pm0.11$\,days) become non-negligible.

Next, we fit a realistic synthetic sample drawn from a uniform spatial distribution out to $z=0.06$. Because the sample is dominated by objects near this outer boundary, the inferred population means closely track the $z=0.06$ results. With a 40\% truncation threshold, we find only a marginal increase in the inferred scatter: $\sigma_{\alpha_g}=0.030\pm0.005$, $\sigma_{\alpha_r}=0.006\pm0.003$, and $\sigma_{t_\mathrm{rise}}=0.017\pm0.008$\,days. These values remain negligible compared to the intrinsic population scatter measured in real data (see Section~\ref{sec:ztf}). Even with a 30\% truncation threshold, the induced scatter remains small ($\sigma_{\alpha_g}=0.025\pm0.006$, $\sigma_{\alpha_r}=0.022\pm0.008$, and $\sigma_{t_\mathrm{rise}}=0.012\pm0.007$\,days). This demonstrates that in a volume-limited sample, the scatter induced by neglecting $K$-corrections is consistently low regardless of the truncation threshold, and any induced parameter correlations are consistent with zero.

In conclusion, while ignoring $K$-corrections introduces systematic shifts in the population mean of certain parameters (most notably $\alpha_g$), the resulting artificial scatter and correlations are negligible relative to the intrinsic distributions and typical measurement uncertainties in current volume-limited SN\,Ia surveys. We note, however, that this test relies entirely on the SED sequence of SN\,2011fe, which belongs to the early-red population \citep{stritzingerRedBlueEarly2018}. Its early spectra are characterized by a cool photosphere with strong absorption in the blue optical (overlapping the ZTF $g$ band). Early-blue SNe\,Ia, which possess hotter photospheres and relatively featureless early spectra, should be less affected by $K$-corrections in the $g$ band. A larger sample of early-time spectrophotometry is needed to fully quantify these effects across the SN\,Ia population.

\section{Population-prior Regularized Fits for Individual ZTF SNe\,Ia} 
\label{sec:ztf_pop_prior}
\begin{figure*}
    \centering
    \includegraphics[width=\linewidth]{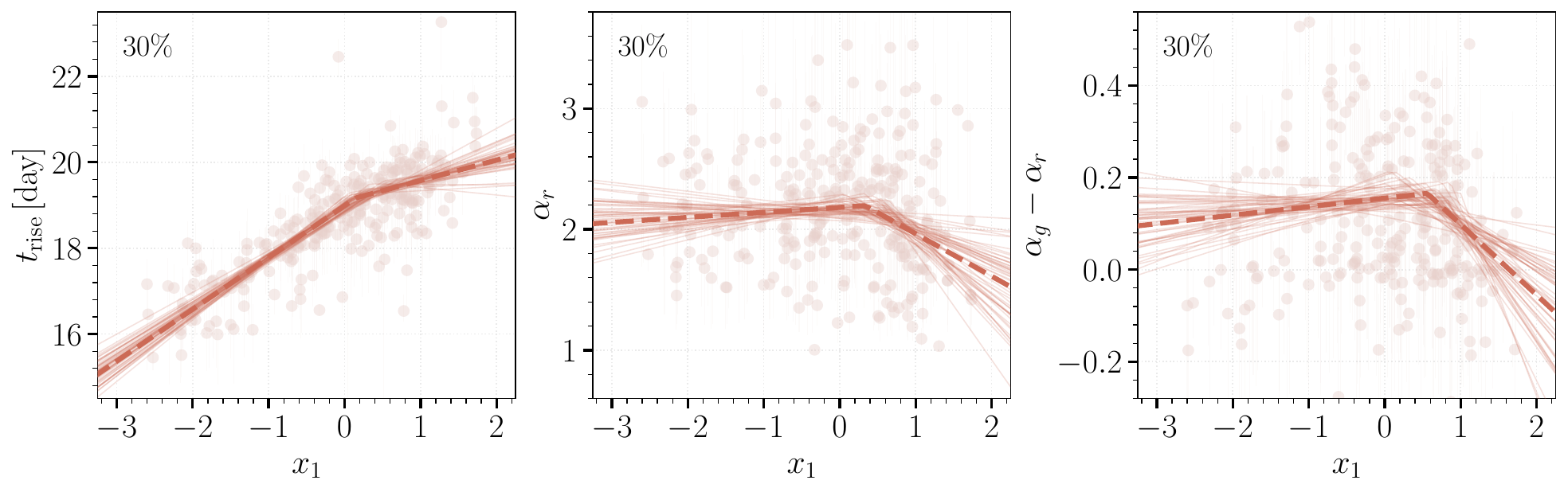}
    \caption{
        Correlations between early-time parameters and the full light-curve stretch $x_1$ for the same subset of 272 SNe\,Ia, where the individual parameter estimates are inferred using a custom population prior that retains the correlation structure for the nuisance parameter $\ln A$ but leaves $t_\mathrm{rise}$ and $\alpha$ uninformative.
        The overall trends remain consistent with those observed in Figure~\ref{fig:ztf_t_rise_x1_corr}, although the scatter is notably larger, particularly among the low-stretch SNe\,Ia. Plot formatting follows Figure~\ref{fig:ztf_t_rise_x1_corr}.
     }
    \label{fig:ztf_pop_prior_fits}
\end{figure*}

In Section~\ref{sec:corr_x1}, we present the correlations between early-time parameters and the light-curve stretch for a subset of 272 SNe\,Ia with confined early-time coverage, where the individual parameter estimates are inferred using the hierarchical Bayesian model.
As we showed in \citetalias{PaperI}, while the hierarchical Bayesian model gives us low-variance estimates for individual SNe\,Ia with ZTF-like data quality, the Bayesian shrinkage biases the individual parameter estimates toward the population mean, particularly for objects in the tails of the distribution. 
On the other hand, traditional unpooled models with uninformative priors can yield less biased estimates for individual SNe\,Ia, but suffer from large uncertainties and variance, which may obscure the underlying population-level trends.

To balance these two extremes, we perform fits for individual SNe\,Ia using a custom prior derived from the hierarchical model. Implementing the full population prior on $t_\mathrm{rise}$, $\alpha$, and $\ln A$ would reintroduce the Bayesian shrinkage effect. Instead, we use a prior that retains the full correlation structure and the population mean and scatter for the nuisance parameter $\ln A$, but leaves the priors on $t_\mathrm{rise}$ and $\alpha$ uninformative. This approach reduces bias in the inferred $\alpha$ and $t_\mathrm{rise}$ for individual SNe\,Ia and provides regularization to prevent unphysical fits \citepalias{PaperI}, at the cost of larger parameter variance compared to applying the full population prior.

Figure~\ref{fig:ztf_pop_prior_fits} displays the resulting fits for the same subset of 272 SNe\,Ia, using a fixed truncation threshold of 30\%, with the inferred values of $t_\mathrm{rise}$, $\alpha_r$, and $\alpha_g - \alpha_r$ plotted against $x_1$. The overall trends remain consistent with those observed in Figure~\ref{fig:ztf_t_rise_x1_corr}, although the scatter is notably larger, particularly among the low-stretch SNe\,Ia. 

\begin{deluxetable}{cccc}
    \tablecaption{Posterior Inference for the Broken Linear Model in the Early-Time Properties (Inferred with the Custom Population Prior) -- $x_1$ Relations.
    \label{tab:broken_pop_params}}
    \tablehead{
    \colhead{Parameter} & \colhead{$t_\mathrm{rise}$\,[day]} & \colhead{$\alpha_r$} & \colhead{$\alpha_g - \alpha_r$}
    }
    \startdata
    $\beta_0$ & $18.99^{+0.08}_{-0.09}$ & $2.18^{+0.06}_{-0.05}$ & $0.16^{+0.02}_{-0.02}$ \\
$\beta_1$ & $1.21^{+0.09}_{-0.09}$ & $0.04^{+0.06}_{-0.06}$ & $0.02^{+0.02}_{-0.02}$ \\
$\beta_2$ & $0.48^{+0.16}_{-0.17}$ & $-0.35^{+0.12}_{-0.15}$ & $-0.15^{+0.06}_{-0.07}$ \\
$\beta_1 - \beta_2$ & $0.74^{+0.21}_{-0.21}$ & $0.40^{+0.16}_{-0.15}$ & $0.17^{+0.07}_{-0.06}$ \\
$x_b$ & $0.15^{+0.16}_{-0.10}$ & $0.34^{+0.29}_{-0.20}$ & $0.56^{+0.21}_{-0.29}$ \\
$\epsilon$ & $0.22^{+0.06}_{-0.06}$ & $0.31^{+0.03}_{-0.03}$ & $0.08^{+0.01}_{-0.01}$ \\
$\nu$ & $2.4^{+0.9}_{-0.6}$ & $18.4^{+15.4}_{-9.2}$ & $18.9^{+16.4}_{-9.1}$ \\
    \enddata
    \end{deluxetable}

Table~\ref{tab:broken_pop_params} details the optimal parameters for the same broken-linear model employed in Section~\ref{sec:corr_x1}. The positions of $x_\mathrm{brk}$ align well with those derived from the hierarchical model estimates. Similarly, the changes in slope across the break, defined by $\beta_1 - \beta_2$, are statistically significant ($\gtrsim$$3\sigma$) for all three parameters, reinforcing the validity of the observed correlations. We note, however, that the slope in the $t_\mathrm{rise}$--$x_1$ relation among the high-stretch population becomes marginally steeper ($\Delta \beta_2 = 0.31\pm0.24$\,days). This implies that Bayesian shrinkage may have slightly attenuated the slope in the hierarchical model fits by pulling the longest risers closer to the population mean. For both $\alpha_r$ and $\alpha_g - \alpha_r$, the slopes for the low-stretch population $\beta_1$ remain statistically consistent with zero, confirming that the absence of correlation in this regime is a genuine feature and not an artifact of Bayesian shrinkage.

\bibliography{sn,facility,ML,cross_ref}{}
\bibliographystyle{aasjournalv7}


\end{document}